\documentclass[conference]{IEEEtran}

\ifCLASSINFOpdf

\else

\fi

\usepackage{wrapfig}
\usepackage{rotating}
\usepackage{lineno,hyperref}
\usepackage{amssymb}
\usepackage{multirow}
\usepackage{caption}
\usepackage{enumerate}
\usepackage{amsmath}
\newtheorem{defn}{Definition}

\usepackage{graphicx}
\usepackage{booktabs}
\usepackage{caption}
\usepackage{algorithm}
\usepackage[noend]{algorithmic}
\usepackage{url}
\usepackage{graphicx}
\usepackage{caption}
\usepackage{subcaption}
\newtheorem{exmp}{Example}[section]
\usepackage{epstopdf}
\usepackage[tracking=true]{microtype}
\usepackage{tikz}

\newcommand{\compresslist}{
\setlength{\itemsep}{1pt}
\setlength{\parskip}{0pt}
\setlength{\parsep}{0pt}
}
\usepackage{eqparbox}

\usepackage{mathtools}

\DeclareMathAlphabet\mathbfcal{OMS}{cmsy}{b}{n}

\newcommand{\RNum}[1]{\uppercase\expandafter{\romannumeral #1\relax}}
\DeclareMathAlphabet\mathbfcal{OMS}{cmsy}{b}{n}

\usepackage{mathtools}

\begin{document}
\title{Improved Solution to the Non-Domination Level Update Problem}

\author{\IEEEauthorblockN{Sumit Mishra}
\IEEEauthorblockA{Department of Computer Science\\ \& Engineering\\
Indian Institute of Technology Patna\\
Patna, Bihar India - 800013\\
sumitmishra@iitp.ac.in}
\and
\IEEEauthorblockN{Samrat Mondal}
\IEEEauthorblockA{Department of Computer Science\\ \& Engineering\\
Indian Institute of Technology Patna\\
Patna, Bihar India - 800013\\
samrat@iitp.ac.in}
\and
\IEEEauthorblockN{Sriparna Saha}
\IEEEauthorblockA{Department of Computer Science\\ \& Engineering\\
Indian Institute of Technology Patna\\
Patna, Bihar India - 800013\\
sriparna@iitp.ac.in}}

\maketitle

\begin{abstract}
Non-domination level update problem is to sort the non-dominated fronts after insertion or deletion of a solution. Generally the solution to this problem requires to perform the complete non-dominated sorting which is too expensive in terms of number of comparisons. Recently an Efficient Non-domination Level Update (ENLU) approach is proposed which does not perform the complete sorting. For this purpose, in this paper a space efficient version of ENLU approach is proposed without compromising the number of comparisons. However this approach does not work satisfactorily in all the cases. So we have also proposed another tree based approach for solving this non-domination level update problem. In case of insertion, the tree based approach always checks for same number of fronts unlike linear approach in which the number of fronts to be checked depends on the inserted solution. The result shows that in case where all the solutions are dominating in nature the maximum number of comparisons using tree based approach is $\mathcal{O}(\log N)$ as opposed to $\mathcal{O}(N)$ in ENLU approach. When all the solutions are equally divided into $K$ fronts such that each solution in a front is dominated by all the solutions in the previous front then the maximum number of comparisons to find a deleted solution in case of tree based approach is $K{-}\log K$ less than that of ENLU approach. Using these approaches an on-line sorting algorithm is also proposed and the competitive analysis of this algorithm is also presented.
\end{abstract}

\IEEEpeerreviewmaketitle

\section{Introduction}
In past few decades the evolutionary algorithms \cite{back1996evolutionary}, \cite{back1993overview}, \cite{back2000evolutionary}, \cite{whitley1996evaluating} have gained a lot of popularity. One of the primarily reason behind its popularity is their ability to solve real world problems. Real world problems may involve simultaneous optimizing multiple objectives. Thus the evolutionary algorithms also optimize single as well as multiple objectives. In case of single objective optimization, only single solution is optimal one. But in case of multi-objective optimization  problems (MOOPs) \cite{deb2001multi}, \cite{coello2002evolutionary}, \cite{deb2000fast}, \cite{zitzler1999evolutionary} a set of set of optimal solutions are achieved and these are known as Pareto-optimal solutions.

In literature various multi-objective evolutionary algorithms (MOEAs) are proposed. Some of them are  non-dominated sorting genetic algorithm II (NSGA-II) \cite{deb2002fast}, strength pareto evolutionary algorithm 2 (SPEA2) \cite{zitzler2001spea2}, pareto archive evolution strategy (PAES) \cite{knowles2000approximating}, pareto envelope-based selection algorithm (PESA) \cite{corne2000pareto} and pareto frontier differential evolution (PDE) \cite{abbass2001pde} etc. These MOEAs are able to find a set of Pareto optimal solutions in one single run. There exist various approaches for the selection of solutions \cite{berry2005combative}. But the Pareto-based approach is generally used. Non-dominated sorting \cite{deb2002fast} is found to be efficient for finding Pareto-optimal solution out of various techniques.

In this sorting the solutions are assigned to their respective front based on their dominance relationship. This process is time consuming when the number of solutions in the populations becomes larger. Much work \cite{deb2002fast}, \cite{mcclymont2012deductive}, \cite{zhang2015efficient} has been done to improve the running time of this process. Golberg et al.~\cite{golberg1989genetic} first proposed the idea of non-dominated sorting. Later this idea was used in multi-objective genetic algorithm \cite{srinivas1994muiltiobjective}.

Let $\mathbb{P} = \left\lbrace p_1,p_2,\ldots,p_N \right\rbrace$ be the population of $N$ solutions. These solutions are categorized into $K$ fronts. These $K$ fronts are denoted as $F_k, 1 \leq k \leq K$. The solutions which belong to front $F_k$ are dominated by at-least one of the solutions belonging to front $F_k'$, $k' < k, k,k' = 1,2,\ldots,K$. Let the number of solutions in each front $F_k$ is $n_k, 1 < k < K$. Thus $N = \sum_{k=1}^{K}n_k$. The arrangement of the solutions in each front is shown in Table \ref{tab:general}. Consider an example. 
\begin{exmp}
Let $\mathbb{P}$ be a set of $12$ solutions. Two objectives are associated with each solution. Let both the objectives are to be minimized. These $12$ solutions are arranged in $5$ fronts. The arrangement of solutions in each fronts is as follows: $F_1 = \left\lbrace p_1 \right\rbrace$, $F_2 = \left\lbrace p_2,p_3 \right\rbrace$, $F_3 = \left\lbrace p_4,p_5,p_6,p_7 \right\rbrace$, $F_4 = \left\lbrace p_8,p_9,p_{10},p_{11} \right\rbrace$, $F_5 = \left\lbrace p_{12} \right\rbrace$.
\end{exmp}

\begin{table}
\centering
\begin{tabular}{|c|c|}
\hline
\textbf{Front}    &  \textbf{Solutions} \\
\hline
$F_1$    &  $sol_1,sol_2,\ldots,sol_{n_1}$ \\ 
$F_2$    &  $sol_{n_1+1},sol_{n_1+2},\ldots,sol_{n_1+n_2}$ \\ 
\vdots   &   \vdots   \\ 
$F_K$    &  $sol_{n_1+\cdots+n_{\textit{K-1}}+1},\ldots,sol_{N}$ \\ 
\hline
\end{tabular}
\caption{Solutions in $K$ different fronts}
\label{tab:general}
\end{table}

The generational Evolutionary Multi-Objective Optimization (EMO) algorithms generate all the offspring solutions from the parent solutions. Then they both are compared. As opposite to the generational EMO algorithms, steady state EMO algorithms \cite{deb2005evaluating}, \cite{beume2007sms} update the parent population as a new offspring is derived. The steady state EMO algorithms have the ability to generate the good offspring solutions. The parallel implementation is also possible with this kind of EMO algorithms.

There have not been so many such kind of EMO algorithms proposed \cite{li2014efficient}. One of the primarily reason for this is the overhead in repeatedly performing the non-dominated sorting as a new solution is generated or an existing solution is removed. But when either a new solution in inserted or an existing solution is removed then not all the solutions change its domination level so it is un-necessary to perform the complete non-dominated sorting again and again. Only some of the solutions need to change their non-domination level. This is first addressed in \cite{li2014efficient}. Li at al.~\cite{li2014efficient} proposed an efficient non-domination level update (ENLU) approach for steady-state EMO algorithms. They have proposed the approach for insertion as well as deletion. In this approach not all the solutions change their domination level. The solutions which need to change their domination level, only change the level. After this some more work has been carried out in this direction \cite{yakupov2015incremental}, \cite{buzdalov2015fast}. But these two work focus on bi-objective steady state EMO algorithms. So in this paper we have proposed the modified version of ENLU which is efficient in terms of space and time complexity remains the same. One more approach is provided which is based on tree data structure. The maximum number of dominance comparison while performing insertion or deletion is also obtained. In short the main contribution are as follows:

\begin{itemize}\compresslist
\item The modified linear approach with space requirement $\mathcal{O}(1)$ as opposite to $\mathcal{O}(N)$ in \cite{li2014efficient} is proposed.
\item The dominance tree based approach is proposed for non-domination level update problem.
\item We have obtained the maximum number of dominance comparisons occurred in linear as well as dominance tree based approach. 
\item The approach for searching a solution is also proposed using dominance tree.
\item The solution to the non-domination level update problem can be used as a non-dominated sorting algorithm. This sorting algorithm can be used as on-line algorithm. The competitive ratio of this on-line algorithm is also obtained.
\end{itemize}

The rest of this paper is organized as follows. Section \ref{sec:enlu} describes the Non-domination Level Update problem.  The related work is also described in this Section. The modified linear approach to insert a solution in the given set of fronts is discussed in Section \ref{sec:mliInsert}. The Proposed dominance binary search tree based approach is illustrated in Section \ref{sec:proposedApproachTree}. The look up procedure using dominance binary search tree based approach is provided in Section \ref{sec:lookup}. The procedure to delete an existing solution is discussed in Section \ref{sec:liDelete}. Section \ref{sec:dominanceComparison} discuss the maximum number of dominance comparison in Non-domination Level Update problem. In section \ref{sec:nd} and \ref{sec:d} we discuss the maximum and minimum number of dominance comparison in two cases when al the solutions are non-dominating and when all are dominating respectively. The sorting algorithm using the proposed approach is presented in Section \ref{sec:sorting}. Finally Section \ref{sec:conclusion} concludes the paper and provides the future direction of the work.

\section{Non-domination Level Update problem}\label{sec:enlu}
In this section we discuss the non-domination level update problem. Non-domination Level Update problem is to sort the non-dominating front after insertion of a new solution or after deletion of an existing solution. Let $\mathbb{P} = \left\lbrace sol_1,sol_2,\ldots,sol_N \right\rbrace$ be the set of $N$ solutions. Let $M$ objectives are associated with these $N$ solutions. These $N$ solutions are divided into $K$ fronts. Let $\mathcal{F} = \left\lbrace F_1,F_2,\ldots,F_{K-1},F_K \right\rbrace$ be the set of $K$ fronts in decreasing order of their dominance i.e., the first front is having rank $1$ (non-domination level $1$), second is having rank $2$ and so on. The number of solutions in each front $F_i$ is given by $n_i$, $1 \leq i \leq K$. Thus $\sum_{i=1}^{K}n_i = N$. Table \ref{tab:general} shows this scenario. 

\begin{figure*}
\centering
\begin{subfigure}{.30\textwidth}
  \centering
  \includegraphics[scale=0.40]{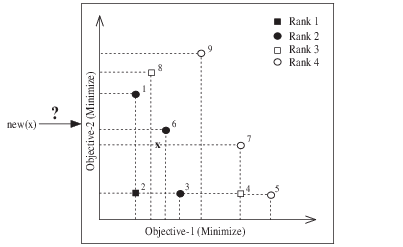}
  \caption{Set of fronts $\mathcal{F}$ before insertion of $\textit{new}$.}
  \label{fig:ndlui1}
\end{subfigure}~~
\begin{subfigure}{.30\textwidth}
  \centering
  \includegraphics[scale=0.40]{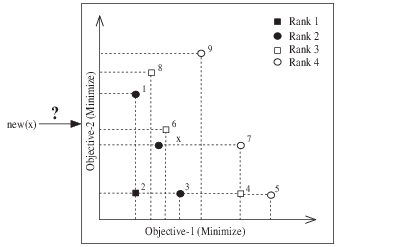}
  \caption{Set of fronts $\mathcal{F}$ after insertion of $\textit{new}$.}
  \label{fig:ndlui2}
\end{subfigure}~~
\begin{subfigure}{.30\textwidth}
  \centering
  \includegraphics[scale=0.40]{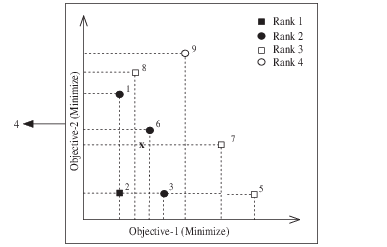}
  \caption{Set of fronts $\mathcal{F}$ after deletion of $4$.}
  \label{fig:ndlud2}
\end{subfigure}
\caption{Schematic diagram of Non-domination Level Update problem in case of insertion.}
\label{fig:ndlui}
\end{figure*}

Let a new solution $\textit{new}$ is to be inserted into the set of $K$ fronts. Non-domination Level Update problem in case of insertion is to insert this new solution at its correct position in the set of fronts and update the fronts if needed. Figure \ref{fig:ndlui1} clearly shows this situation. In this figure, there are $9$ solutions. Let $\mathbb{P} = \left\lbrace 1,2,\ldots,9 \right\rbrace$ be the set of these $9$ solutions. These solutions are divided into $4$ fronts. Let $\mathcal{F} = \left\lbrace F_1,F_2,F_3,F_4 \right\rbrace$ be the set of $4$ fronts in the decreasing order of their dominance. $F_1 = \left\lbrace 2 \right\rbrace$, $F_2 = \left\lbrace 1,3,6 \right\rbrace$, $F_3 = \left\lbrace 4,8 \right\rbrace$, $F_4 = \left\lbrace 5,7,9 \right\rbrace$. A new solution $\textit{new}$ is to be inserted in the set of fronts. Main aim here is to update $\mathcal{F}$ after insertion of $\textit{new}$. Figure \ref{fig:ndlui2} shows the updated set of fronts when new solution is inserted in the set of fronts.

Let an existing solution $sol$ is to be deleted from the set of $K$ fronts. Non-domination Level Update problem in case of deletion is to delete this solution and update the set of fronts if required. Figure \ref{fig:ndlui1} shows the set of $4$ fronts. An existing solution $4$ is to be deleted from the set of fronts. The motive is to update $\mathcal{F}$ after deletion of solution $4$. Figure \ref{fig:ndlud2} shows the updated set of fronts when solution $4$ is being deleted from the set of fronts.

The naive approach is to apply the non-dominated sorting algorithm on all the $N$ solutions along with new solution in case of insertion or apply the non-dominated sorting algorithm on the remaining $N{-}1$ solutions in case of deletion. Thus the complete sorting algorithm is to be applied on either $N{+}1$ or $N{-}1$ solutions. If we use the brute-force technique then the time complexity will be $\mathcal{O}(MN^3)$ and space requirement will be $\mathcal{O}(N)$. When the fast non-dominating sorting proposed in \cite{deb2002fast} is used then the time complexity will be $\mathcal{O}(MN^2)$. This always requires $(N{+}1)N$ dominance comparison in case of insertion and $(N{-}1)(N{-}2)$ comparison in case of deletion regardless of arrangement of fronts. The space requirement will be $\mathcal{O}(N^2)$. Tang et al.~\cite{tang2008fast} proposed arena's principle. The time complexity of their approach is $\mathcal{O}(MN^2)$ in worst case. The space requirement is $\mathcal{O}(N)$. McClymont et al.~\cite{mcclymont2012deductive} proposed two sorting algorithms - climbing sort and deductive sort. The climbing sort requires $(N{+}1)N$ comparison in case of insertion and $(N{-}1)(N{-}2)$ comparison in case of deletion when the worst case scenario occurs. The space requirement is $\mathcal{O}(N^2)$. The deductive sort has some advantage over climbing. In the worst case the number of comparison is $\frac{(N{+}1)N}{2}$ for insertion and $\frac{(N{-}1)(N{-}2)}{2}$ for deletion. The space requirement is $\mathcal{O}(N)$ as opposite to $\mathcal{O}(N^2)$ by climbing sort. The approach proposed in \cite{zhang2015efficient} requires $\mathcal{O}(MN^2)$ time in worst case. In the worst case the total number of dominance comparison is same as deductive sort. This approach first sorts the solutions based on first objective which also requires $\mathcal{O}(N\log N)$ time. The space requirement of this approach is $\mathcal{O}(1)$.

All the above discussed approaches apply the complete sorting algorithm on either $N{+}1$ or $N{-}1$ solutions. To reduce the number of dominance comparisons without applying the complete sorting algorithm, Li et al.~\cite{li2014efficient} proposed an approach for Non-domination Level Update problem. The approach was proposed for Steady-State Evolutionary Multiobjective Optimization. The time complexity is $\mathcal{O}(MN\sqrt{N})$ in case of equal division of solutions in $\sqrt{N}$ fronts. However the worst case time complexity is $\mathcal{O}(MN^2)$ \cite{yakupov2015incremental}, \cite{buzdalov2015fast}. The space requirement is $\mathcal{O}(N)$ as the dominated solution is kept at tow places - one is in archive $S$ and other is in front $F_{i{+}1}$.

Yakupov et al.~\cite{yakupov2015incremental} proposed an incremental non-dominated sorting for bi-objective solutions. The running time of the approach is $\mathcal{O}(M(1+\log(N/M))+ \log M \log (N/\log M))$ which is $\mathcal{O}(N)$ in worst case when a new solution is inserted in the set of fronts having $N$ solutions with $M$ objectives. With further improvement in this work, Buzdalov et al.~\cite{buzdalov2015fast} proposed fast implementation of the steady-State NSGA-II algorithm for two dimensions based on incremental non-dominated sorting. In this work, the support of the crowding
distance calculation is employed. After that, the steady-state version of the NSGA-II algorithm is presented. In \cite{yakupov2015incremental} and \cite{buzdalov2015fast} a data structure tree of tress is used. The nodes in higher level tree stores the number of tree elements in a sub-tree, the previous-in-order node and the next-in-order node. Thus extra space is required which is $\mathcal{O}(K)$. The nodes in lower level tree also stores the number of tree elements in a sub-tree, the previous-in-order node and the next-in-order node. Thus extra space is required which is $\mathcal{O}(N)$. Thus the space requirement is $\mathcal{O}(K) + \mathcal{O}(N) \equiv \mathcal{O}(N)$. 

We have proposed an approach based on dominance binary search tree. First we discuss the modified version of linear approach proposed in \cite{li2014efficient} for insertion as well as deletion. The modification is done to reduce the space complexity from $\mathcal{O}(N)$ to $\mathcal{O}(1)$.

\section{Modified Linear Approach: Insert a Solution}\label{sec:mliInsert}
In this section we will discuss the Non-domination Level Update problem using linear way to obtain the correct position of new solution $\textit{new}$ as done in \cite{li2014efficient}. Algorithm \ref{alg:InsertLinear} shows the insertion procedure of $\textit{new}$ in the list of non-dominated fronts $\mathcal{F}$. This algorithm does not run the complete sorting algorithm again. It uses the non-dominance properties of the solution in the same front and uses the ranking of the front. Here we are assuming that all the fronts are arranged in decreasing order of their dominance. The process for inserting a solution is summarized in Algorithm \ref{alg:InsertLinear}. Two solutions are compared for dominance nature using \textit{domNature(A,B)} procedure which returns the following three values.
\begin{itemize}\compresslist
\item 1: Solution $A$ dominates $B$.
\item -1: Solution $A$ is dominated by $B$.
\item 0: Solution $A$ and $B$ are non-dominated.
\end{itemize}

\begin{algorithm}
\begin{algorithmic}[1]
\renewcommand{\algorithmicrequire}{\textbf{Input:}}
\REQUIRE  $\mathcal{F} = \left\lbrace F_1,F_2,\ldots,F_K \right\rbrace$: Non-dominated fronts in the decreasing order of their dominance\\ 
\hspace{3.08 cm} $\textit{new}$: A new solution 
\renewcommand{\algorithmicensure}{\textbf{Output:}}
\ENSURE Updated $\mathcal{F}$ after insertion of $\textit{new}$
\FOR{$i \leftarrow 1 \text{ to } K$}
	\STATE $\textit{count} \leftarrow 0$ 
	\STATE $\textit{new}_{\textit{dom}} \leftarrow \Phi$
	\FOR{$j \leftarrow 1 \text{ to } \vert F_i \vert$}
		\STATE  $\textit{isdom} \leftarrow$ {\textit{domNature}($\textit{new},F_i(j)$)} 
		\IF{$\textit{isdom} = 1$}
			\STATE $\textit{new}_{\textit{dom}} \leftarrow \textit{new}_{\textit{dom}} \cup \left\lbrace F_i(j) \right\rbrace$
			\STATE $F_i \leftarrow F_i \setminus \left\lbrace F_i(j) \right\rbrace$
			\STATE $\textit{DomSet}(F_i,\textit{new},i,\textit{new}_{\textit{dom}})$							
			\STATE $F_i \leftarrow F_i \cup \textit{new}$	
			\IF{$i = K$}
				\STATE Make $\textit{new}_{\textit{dom}}$ a new front $F_{K{+}1}$
			\ELSIF{$\vert F_i \vert = 1$}
				\STATE Increase the dominance level of all the fronts $F_{k+1}, F_{k+2}, \ldots, F_K$ by $1$ and make $\textit{new}_{\textit{dom}}$ a new front $F_{k+1}$
			\ELSE			
				\STATE \textit{UpdateInsert}$(\mathcal{F},\textit{new}_{\textit{dom}},i{+}1)$
			\ENDIF
			\RETURN $\mathcal{F}$
		\ELSIF{$\textit{isdom} = 0$}
			\STATE	$\textit{count} \leftarrow \textit{count}+1$
		\ELSE%[check for next front]
			\STATE break  
		\ENDIF	
	\ENDFOR
	\IF{$\textit{count} = \vert F_i \vert$}
		\STATE Insert $\textit{new}$ in $F_i$  \hfill /* new is non-dominated with $F_i$
		\RETURN $\mathcal{F}$
	\ENDIF
\ENDFOR
\STATE Make $\textit{new}$ a new front $F_{K+1}$
\RETURN $\mathcal{F}$
\end{algorithmic}
\caption{InsertLinear$(\mathcal{F},\textit{new})$}
\label{alg:InsertLinear}
\end{algorithm}

\begin{algorithm}
\begin{algorithmic}[1]
\renewcommand{\algorithmicrequire}{\textbf{Input:}}
\REQUIRE $F$: A non-dominated front \\
\hspace{0.13 cm} $\textit{new}$: A new solution \\
$\textit{index}$: Index of the solution from where the dominance need to be checked \\
\hspace{0.40 cm} $S$: Set of solutions dominated by $\textit{new}$
\renewcommand{\algorithmicensure}{\textbf{Output:}}
\ENSURE Updated $S$
\FOR{$i \leftarrow \textit{index} \text{ to } \vert F \vert$}
	\STATE $\textit{isdom} \leftarrow$ {domNature($\textit{new},F(i)$)} 
	\IF{$\textit{isdom} = 1$}
		\STATE $S \leftarrow S \cup \left\lbrace F(i) \right\rbrace$
		\STATE $F \leftarrow F \setminus F(i)$
		\STATE $i \leftarrow i{-}1$
	\ENDIF	
\ENDFOR 
\RETURN $S$
\end{algorithmic}
\caption{DomSet$(F,\textit{new},\textit{index},S)$}
\label{alg:DomSet}
\end{algorithm}

The new solution $\textit{new}$ is compared with all the fronts in sequential manner starting from $\mathcal{F}_1$ to $\mathcal{F}_K$. $\textit{new}$ is compared with each solution in a front. When $\textit{new}$ is compared with any solution in a front $F_k (1 \leq k \leq K)$ then there are three possibilities: 

\noindent
1. If the solution in the front $F_k$ dominates $\textit{new}$ it means $\textit{new}$ can not be inserted into $F_k$. So now we check for another front having lower dominance $(F_{k+1})$ without checking it with other solutions in the same front. If $i=K$ then $\textit{new}$ creates a new front $F_{K+1}$ having lowest dominance among all the fronts.

\noindent
2. If the solution in the front is non-dominating with the $\textit{new}$ then we keep on comparing $\textit{new}$ with rest of the solutions in that front. If $\textit{new}$ is non-dominating with all the solutions in the front $F_k$ then $\textit{new}$ is added in the front $F_k$.

\noindent
3. If $\textit{new}$ dominates the solution in the front $F_k$ then we obtain the list $\textit{new}_{\textit{dom}}$ of all the solutions in the front $F_k$ which are dominated by $\textit{new}$ using \textit{DomSet()} procedure as described in Algorithm \ref{alg:DomSet}. This \textit{DomSet()} procedure returns the list of all the solutions in front $F_k$ which are dominated by $\textit{new}$. The solutions which are dominated by $\textit{new}$ are removed from front $F_k$. After the removal of dominated solution in $F_k$, $\textit{new}$ is added to $F_k$. Now we have the following possibilities:

\begin{itemize}\compresslist
\item If $k=K$ then $\textit{new}_{\textit{dom}}$ creates a new front $F_{K+1}$ having lowest dominance among all the fronts.
\item If $\textit{new}$ dominates all the solutions in the front (after removing the dominated solutions from $F_k$ and adding $\textit{new}$ to $F_k$ the cardinality of $F_k$ is $1$) i.e. $\vert F_k \vert = 1$ then the dominance level of all the fronts $F_{k+1}, F_{k+2}, \ldots, F_K$ are increased by one and dominated solution $\textit{new}_{\textit{dom}}$ is assigned to front $F_{k+1}$.
\item If none of the above two conditions are satisfied i.e. $\textit{new}$ dominates some of the solutions in front $F_k$ and $k<K$ then \textit{UpdateInsert()} procedure is used which re-arranges the solutions in their respective front. This \textit{UpdateInsert()} procedure is discussed in Algorithm \ref{alg:UpdateInsert}.
\end{itemize}

\subsection{Illustration of DomSet$(F,\textit{new},\textit{index},S)$ procedure}
This procedure takes as input a front $(F)$, new solution $(\textit{new})$, the index of the solution $(\textit{index})$ in the front from where the dominance of $\textit{new}$ needs to be checked and set of solutions dominated by $\textit{new}$ $(S)$. This procedure returns the updated set of solutions dominated by $\textit{new}$ in front $F$. For this purpose $\textit{new}$ is checked against all the solutions in the front starting from index position. The solution which is dominated by $\textit{new}$ is added to $S$ and removed from the front $F$. We are removing the dominated solutions from the front to save the space. In this manner same solution does not occupy more than one place.

\noindent
\textbf{Complexity Analysis:}
In the worst case, all the solutions in the front except first is compared with new solution in the \textit{DomSet()} procedure so the worst case complexity of \textit{DomSet()} procedure becomes $\mathcal{O}(M\vert F \vert)$.

\begin{algorithm}
\begin{algorithmic}[1]
\renewcommand{\algorithmicrequire}{\textbf{Input:}}
\REQUIRE $\mathcal{F}$: Set of non-dominated fronts \\ 
\hspace{0.0001 cm} $\textit{index}$: Non-domination level index
\renewcommand{\algorithmicensure}{\textbf{Output:}}
\ENSURE Updated $\mathcal{F}$ 
	\STATE $l \leftarrow \vert S \vert$
	\FOR{$i \leftarrow 1 \text{ to } \vert F_{\textit{index}} \vert$}
		\STATE $\textit{count} \leftarrow 0$
		\FOR{$j \leftarrow 1 \text{ to } l$}
			\IF{$\textit{domNature}(S(j), F_{\textit{index}}(i)) = 0$}
				\STATE $\textit{count} \leftarrow \textit{count} {+} 1$
			\ENDIF
		\ENDFOR
		\IF{$\textit{count} = l$}
			\STATE $S \leftarrow S \cup F_{\textit{index}}(i)$
			\STATE $F_{\textit{index}} \leftarrow F_{\textit{index}} \setminus F_{\textit{index}}(i)$ 
			\STATE $i \leftarrow i - 1$
		\ENDIF
	\ENDFOR
	\IF{$l = \vert S \vert$}
		\STATE Increase the domination level of $F_k, k \in \left\lbrace \textit{index},\textit{index}{+}1,\ldots,K \right\rbrace$ by $1$ and make $S$ a new front $F_{\textit{index}}$
	\ELSIF{$F_{\textit{index}} = \Phi$}
		\STATE Make $S$ as $F_{\textit{index}}$
	\ELSE
		\STATE $T \leftarrow F_{\textit{index}}$ \hfill /*Move the solutions from $F_{\textit{index}}$ to $T$
		\STATE Make $S$ as $F_{\textit{index}}$
		\STATE $\textit{UpdateInsert}(\mathcal{F},T,\textit{index}{+}1)$
	\ENDIF
\end{algorithmic}
\caption{UpdateInsert$(\mathcal{F},S,\textit{index})$}
\label{alg:UpdateInsert}
\end{algorithm}

\subsection{Illustration of UpdateInsert$(\mathcal{F},S,\textit{index})$ procedure}
This procedure takes as input the set of non-dominated fronts $\mathcal{F}$, a set of solution $S$, the index of the front denoted by $\textit{index}$ in $\mathcal{F}$. This procedure updates the $\mathcal{F}$ by either creating a new front or by re-arranging the solutions in the existing fronts. First of all the initial cardinality of $S$ is stored is $l$. This is because when the solutions from $F_{\textit{index}}$ is compared with $S$ for non-dominance then it should be compared with only first $l$ solutions. 

Here we find the solutions in $F_{\textit{index}}$ which are non-dominated with $S$. The solutions which are non-dominated with $S$ are added to it and removed from $F_{\textit{index}}$. The removal guarantees that no solution can occupy more than one place. Now the following situation can occur:

\begin{itemize}\compresslist
\item If no solution from $F_{\textit{index}}$ is being added to $S$ i.e. $l = \vert S \vert$ then the dominance level of fronts $F_k, k \in \left\lbrace \textit{index},\textit{index}{+}1,\ldots,K \right\rbrace$ is increased by $1$ and $S$ is assigned the dominance level $\textit{index}$. 
\item If all the solutions from $F_{\textit{index}}$ is added to $S$ i.e. $F_{\textit{index}} = \Phi$ then make $S$ as $F_{\textit{index}}$.
\item Otherwise all the solutions from $F_{\textit{index}}$ is moved to $T$. The movement means as a solution from  $F_{\textit{index}}$ is moved to $T$, the solution is being removed from $F_{\textit{index}}$. The non-domination level of $F_{\textit{index}}$ is assigned to $S$. The procedure is repeated with UpdateInsert$(\mathcal{F},T,\textit{index}{+}1)$.
\end{itemize}

\noindent
\textbf{Complexity Analysis:}
In this algorithm the maximum number of comparison is performed when $\textit{new}$ dominates the solutions in the first front $F_1$ i.e. this procedure is called with $\textit{index}=2$. For maximum number of comparison, the $\vert \textit{new}_{\textit{dom}} \vert = n_1{-}1$. The maximum number of comparison occurs when each call to this procedure shifts one solution in higher level front after comparing with all the solutions. Thus the maximum number of comparison for this procedure is given by  $\left( n_1{-}1 \right)n_2+\left( n_2{-}1 \right)n_3+\ldots+\left( n_{k-1}{-}1 \right)n_K$. Each comparison between two solutions requires at-most $M$ comparison between $M$ objectives. Thus the worst case complexity of this procedure is $\mathcal{O}(MN^2)$. The best case occurs when $F_1$ has single solution and $new$ is non-dominating with this solution. In this case only one comparison is required so the best case complexity is $\mathcal{O}(M)$.

\subsection{Complexity of Modified Linear Approach}
Here we will analyze the complexity using the linear approach. We can see from all the algorithms \ref{alg:InsertLinear}, \ref{alg:DomSet} and \ref{alg:UpdateInsert} that no solution is being kept at more than one place. Only few scaler variables are required. Therefore, the space complexity of linear approach is $\mathcal{O}(1)$. The worst case time complexity of \textit{UpdateInsert()} procedure dominates the worst case time complexity of \textit{DomSet()} procedure. The complexity of \textit{UpdateInsert()} procedure is quadratic while \textit{DomSet()} has linear complexity. Thus the overall worst case complexity of the modified approach is $\mathcal{O}(MN^2)$. This time complexity is same as proposed in \cite{li2014efficient}.

\section{Proposed Approach using Dominance Binary Search Tree}\label{sec:proposedApproachTree}
In this section we will discuss the proposed approach for Non-domination Level Update problem using dominance tree. This approach inserts the new solution to its correct position using tree based methodology unlike linear way as proposed in \cite{li2014efficient}. The only difference between the linear approach and tree based approach is in identifying the position of the inserted solution. The update procedure remains the same. Thus dominance tree based approach will be beneficial when there are large number of fronts. We first provide formal definition of this tree. Two variants of this tree are also discussed. 

\begin{figure*}
        \centering
        \begin{subfigure}[b]{0.3\textwidth}
                \includegraphics[scale=0.3]{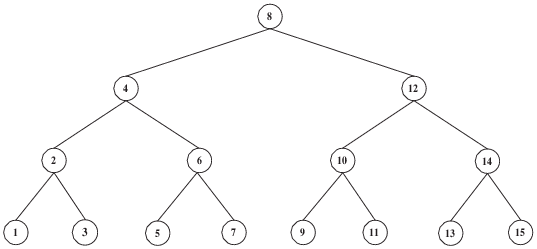}
                \caption{Dominance Binary Search Tree}
                \label{fig:dbst}
        \end{subfigure}
        ~ %add desired spacing between images, e. g. ~, \quad, \qquad, \hfill etc.
        \begin{subfigure}[b]{0.3\textwidth}
                \includegraphics[scale=0.3]{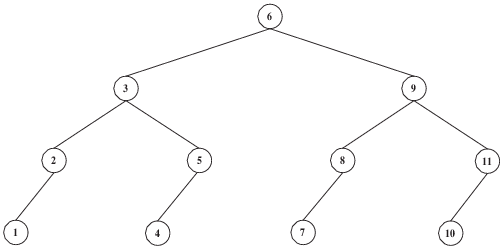}
  				\caption{Left-Balanced Dominance BST}
                \label{fig:dbstleft}
        \end{subfigure}
        ~ %add desired spacing between images, e. g. ~, \quad, \qquad, \hfill etc.
        \begin{subfigure}[b]{0.3\textwidth}
               \includegraphics[scale=0.3]{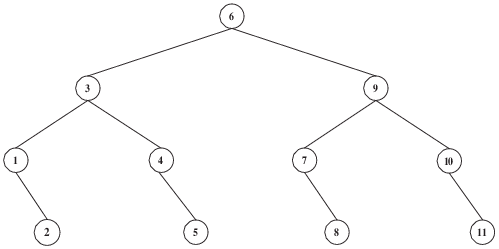}
  				\caption{Right-Balanced Dominance BST}
                 \label{fig:dbstright}
        \end{subfigure}
        \caption{Dominance Binary Search Tree}\label{fig:animals}
\end{figure*}

The tree based data structure is also used in \cite{yakupov2015incremental} and \cite{buzdalov2015fast}. The authors have used tree of trees where the high level tree corresponds to the non-dominated front i.e. node in the high level tree corresponds to a front. The low level tree corresponds to the solutions inside a front i.e. the nodes in the low level tree corresponds to solutions inside the front. The solutions in the low level tree are sorted according to first objective. The high-level tree can be an ordinary balanced tree while the low level tree should be a split-merge tree. The uses of this data structure perform some of the operations in $\mathcal{O}(\log N)$ time like element search in the container, splitting of container by key into two parts and merging of two container. In \cite{buzdalov2015fast}, Cartesian Tree \cite{vuillemin1980unifying} is used as a split-merge tree as it performs better than the Splay Tree \cite{sleator1985self} in practice.

We have used only high level tree named as Dominance Binary Search Tree. The solutions inside a node are arranged in linear fashion. Thus no sorting is required.

\subsection{Dominance Binary Search Tree}
In \textit{Dominance Binary Search Tree}, the node represents single non-dominated front i.e., the solutions in a node are non-dominating with each other.

\begin{defn}[Dominance Binary Search Tree]
A binary tree $T$ is known as \textit{Dominance Binary Search Tree} if the node is having lower dominance than left sub-tree and higher dominance than right sub-tree.
\end{defn}

\begin{exmp}
\textit{Let $\mathcal{F} = \left\lbrace F_1,F_2,\ldots,F_{15} \right\rbrace$ be the set of $15$ non-dominated fronts in decreasing order of their dominance i.e., the first front is having rank 1 (higher rank), second is having rank $2$ and so on. The corresponding \textit{Dominance Binary Search Tree} is given in Figure \ref{fig:dbst}.}
\end{exmp}

Dominance binary search tree can be categorized into two types based on how the root of a sub-tree is chosen. 

\begin{enumerate}
\item \textbf{Left-Balanced Dominance Binary Search Tree:} If left child of each node is filled before the right child then the tree is called left-balanced dominance binary search tree. It is obtained when the index of the root of a sub-tree is calculated as $\textit{mid} = \lceil (\textit{min}+\textit{max})/2 \rceil$. See Figure \ref{fig:dbstleft}.
\item \textbf{Right-Balanced Dominance Binary Search Tree:} If right child of each node is filled before the left child then the tree is called right-balanced dominance binary search tree. It is obtained when the index of the root of a sub-tree is calculated as $\textit{mid} = \lfloor (\textit{min}+\textit{max})/2 \rfloor$. See Figure \ref{fig:dbstright}.
\end{enumerate}

Unlike the linear approach where $\textit{new}$ needs to be compared with all the $K$ fronts in the worst case, here the solution is compared with either $\lfloor \log_2K \rfloor + 1$ or $\lfloor \log_2K \rfloor$ fronts in all the cases. The process to insert a new solution $\textit{new}$ in the set of fronts $\mathcal{F}$ is summarized in Algorithm \ref{alg:InsertTree}.

\begin{algorithm}
\begin{algorithmic}[1]
\renewcommand{\algorithmicrequire}{\textbf{Input:}}
\REQUIRE  $\mathcal{F} = \left\lbrace F_1,F_2,\ldots,F_K \right\rbrace$: Non-dominated front in the decreasing order of their dominance\\ 
\hspace{3.16 cm} $\textit{new}$: A new solution 
\renewcommand{\algorithmicensure}{\textbf{Output:}}
\ENSURE Updated $\mathcal{F}$ after insertion of $new$
\IF{$\vert \mathcal{F} \vert = 1$}
	\STATE $\text{Insert}(\mathcal{F},\textit{new})$ \hfill /* Same as linear approach
\ELSE
	\STATE \textit{CmpFront}[]$<$dom,fIndex,sIndex$>$  $\leftarrow \Phi$
	\STATE InsertTree\_Left$(\mathcal{F},1,\vert \mathcal{F} \vert,\textit{new})$
	\STATE $\textit{len} \leftarrow \textit{CmpFront.size()}$ %\hfill /* Size of the \textit{CmpFront}
		\IF{$\textit{CmpFront}[\textit{len}]_{\text{dom}} = 0$}
			\STATE Insert $\textit{new}$ in $F_{\textit{CmpFront}[\textit{len}]_{\text{fIndex}}}$
		\ELSIF{$\textit{CmpFront}[\textit{len}]_{\text{dom}} = 1$}
			\STATE $\textit{new}_{\textit{dom}} \leftarrow \Phi$
			\STATE $\textit{fIndex} \leftarrow \textit{CmpFront}[\textit{len}]_{\text{fIndex}}$
			\STATE $\textit{sIndex} \leftarrow \textit{CmpFront}[\textit{len}]_{\text{sIndex}}$	
			\STATE $\textit{new}_{\textit{dom}}  \leftarrow  \textit{new}_{\textit{dom}} \cup F_{\textit{fIndex}}(\textit{sIndex})$		
			\STATE $F_{\textit{fIndex}}  \leftarrow   F_{\textit{fIndex}}  \setminus   F_{\textit{fIndex}}(\textit{sIndex})$			
			\STATE DomSet$(F_{\textit{fIndex}},\textit{new},\textit{sIndex}{+}1,\textit{new}_{\textit{dom}})$	
			\STATE $F_{\textit{fIndex}} \leftarrow F_{\textit{fIndex}} \cup \textit{new}$
			
			\IF{$\textit{fIndex} = K$}
				\STATE Make $\textit{new}_{\textit{dom}}$ a new front $F_{\textit{fIndex}{+}1}$
			\ELSIF{$\vert F_{\textit{fIndex}} \vert = 1$}
				\STATE Increase the dominance level of all fronts $F_{\textit{fIndex}{+}1},F_{\textit{fIndex}{+}2},\ldots,K$ by $1$ and make $\textit{new}_{\textit{dom}}$ a new front $F_{\textit{fIndex}{+}1}$
			\ELSE
				\STATE Update$(\mathcal{F},\textit{new}_{\textit{dom}},\textit{fIndex}{+}1)$
			\ENDIF				
		\ELSIF{All the dom value in CmpFront is -1}
			\STATE Make $\textit{new}_{\textit{dom}}$ a new front $F_{K{+}1}$				
		\ELSE
			\FOR{$i \leftarrow \textit{len} \text{ to } 2$}
				\IF{$\textit{CmpFront}[i]_{\text{dom}} \neq \textit{CmpFront}[\textit{i-}1]_{\text{dom}}$}
				\STATE $\textit{fIndex} \leftarrow \textit{CmpFront}[\textit{i-1}]_{\text{fIndex}}$
				\STATE $\textit{sIndex} \leftarrow \textit{CmpFront}[\textit{i-1}]_{\text{sIndex}}$	
				\IF{$\textit{CmpFront}[\textit{i-}1]_{\text{dom}} = 0$}
					\STATE Insert $\textit{new}$ in $F_{\textit{fIndex}}$
				\ELSIF{$\textit{CmpFront}[\textit{i-}1]_{\text{dom}} = 1$}
					\STATE $\textit{new}_{\textit{dom}} \leftarrow \Phi$	
					\STATE $\textit{new}_{\textit{dom}} \leftarrow \textit{new}_{\textit{dom}} \cup F_{\textit{fIndex}}(\textit{sIndex})$	
					\STATE $F_{\textit{fIndex}}  \leftarrow   F_{\textit{fIndex}}  \setminus   F_{\textit{fIndex}}(\textit{sIndex})$	
					\STATE DomSet$(F_{\textit{fIndex}},\textit{new},\textit{sIndex+1},\textit{new}_{\textit{dom}})$
					\STATE UpdateInsert$(\mathcal{F},\textit{new}_{\textit{dom}},\textit{fIndex})$
				\ENDIF	
			\ENDIF
		\ENDFOR
	\ENDIF
\ENDIF
\end{algorithmic}
\caption{InsertTree$(\mathcal{F},\textit{new})$}
\label{alg:InsertTree}
\end{algorithm}

\begin{algorithm*}
\begin{algorithmic}[1]
\renewcommand{\algorithmicrequire}{\textbf{Input:}}
\REQUIRE  $\mathcal{F} = \left\lbrace F_1,F_2,\ldots,F_K \right\rbrace$: Non-dominated front in the decreasing order of their dominance\\ 
\hspace{3.11 cm} $\textit{new}$: A new solution 
\renewcommand{\algorithmicensure}{\textbf{Output:}}
\ENSURE $\textit{CmpFront}$ 
\STATE $\textit{count} \leftarrow 0$
\IF{$\textit{min} = \textit{max}$}
	\FOR{$i \leftarrow 1 \text{ to } \vert F_{\textit{min}} \vert$}
		\STATE $\textit{isdom} \leftarrow$ {domNature$\left( \textit{new},F_{\textit{min}}(i) \right)$}  \COMMENT{check the dominating nature of $\textit{new}$ and $F_{\textit{min}}(i)$}
		\IF[$\textit{new}$ dominates $F_{\textit{min}}(i)$]{$\textit{isdom} = 1$}
			\STATE \textit{CmpFront}.add$(1,\textit{min},i)$\COMMENT{Add dominating nature of $\textit{new}$, front index and solution index}
			\STATE break
		\ELSIF[$\textit{new}$ is dominated by $F_{\textit{min}}(i)$]{$\textit{isdom} = -1$}	
			\STATE \textit{CmpFront}.add$(-1,\textit{min},i)$\COMMENT{Add dominating nature of $\textit{new}$, front index and solution index}
			\STATE break
		\ELSE[$\textit{new}$ and $F_{\textit{min}}(i)$ are non-dominating]
			\STATE $\textit{count} \leftarrow \textit{count}{+}1$
		\ENDIF	
	\ENDFOR
	\IF{$\textit{count} = \vert F_{\textit{min}} \vert$}
		\STATE \textit{CmpFront}.add$(0,\textit{min},0)$\COMMENT{Add dominating nature of $\textit{new}$, front index and solution index}
	\ENDIF
\ELSE
	\STATE $\textit{mid} \leftarrow \lceil \left( \textit{min} + \textit{max} \right) / 2 \rceil$	  \COMMENT{Obtain the position of the node to be explored}
	\FOR{$i \leftarrow \text{ to } \vert F_{\textit{mid}} \vert$}
		\STATE $\textit{isdom} \leftarrow${domNature$\left(\textit{new},F_{\textit{mid}}(i) \right)$}    \COMMENT{check the dominating nature of $\textit{new}$ and $\textit{sol}$}		
		\IF[$\textit{new}$ dominates $\textit{sol}$]{$\textit{isdom} = 1$}		
			\STATE \textit{CmpFront}.add$(\textit{isdom},\textit{mid},i)$\COMMENT{Add dominating nature of $\textit{new}$, front index and solution index}
		    \STATE $\text{InsertTree\_Left}(\mathcal{F},\textit{min},\textit{mid}{-}1,\textit{new})$ \COMMENT{Explore left sub-tree}
			\STATE break
		\ELSIF[$F_{\textit{mid}}(i)$ \textit{dominates} $\textit{new}$]{$\textit{isdom} = -1$}
			\STATE \textit{CmpFront}.add$(\textit{isdom},\textit{mid},i)$\COMMENT{Add dominating nature of $\textit{new}$, front index and solution index}
			\IF[Check for the existence of right sub-tree]{$\textit{mid} \neq \textit{max}$}
				\STATE $\text{InsertTree\_Left}(\mathcal{F},\textit{mid}{+}1,\textit{max},\textit{new})$\COMMENT{Explore right sub-tree}
			\ENDIF
			\STATE break
		\ELSE[$\textit{new}$ and $F_{\textit{mid}}(i)$ are non-dominating]
			\STATE $\textit{count} \leftarrow \textit{count+1}$
		\ENDIF	
	\ENDFOR
	\IF[All solutions in $F_{\textit{\textit{mid}}}$ are non-dominating with $\textit{new}$]{$\textit{count} = \vert F_{\textit{mid}} \vert$}
		\STATE \textit{CmpFront}.add$(0,\textit{mid},0)$\COMMENT{Add dominating nature of $\textit{new}$, front index and solution index}
		\STATE $\text{InsertTree\_Left}(\mathcal{F},\textit{min},\textit{mid}{-}1,\textit{new})$\COMMENT{Explore left sub-tree}
	\ENDIF
\ENDIF
\end{algorithmic}
\caption{InsertTree\_Left$(\mathcal{F},\textit{min},\textit{max},\textit{new})$}
\label{alg:InsertTreeLeft}
\end{algorithm*}

\begin{algorithm*}
\begin{algorithmic}[1]
\renewcommand{\algorithmicrequire}{\textbf{Input:}}
\REQUIRE  $\mathcal{F} = \left\lbrace F_1,F_2,\ldots,F_K \right\rbrace$: Non-dominated front in the decreasing order of their dominance\\ 
\hspace{3.12 cm} $\textit{new}$: A new solution 
\renewcommand{\algorithmicensure}{\textbf{Output:}}
\ENSURE $\textit{CmpFront}$ 
\STATE $\textit{count} \leftarrow 0$
\IF{$\textit{min} = \textit{max}$}
	\FOR{$i \leftarrow \text{ to } \vert F_{\textit{min}} \vert$}
		\STATE $\textit{isdom} \leftarrow$ {dominates$\left( \textit{new},F_{\textit{min}}(i) \right)$}  \COMMENT{check the dominating nature of $\textit{new}$ and $F_{\textit{min}}(i)$}
		\IF[$\textit{new}$ domNature $\textit{sol}$]{$F_{\textit{min}}(i) = 1$}
			\STATE \textit{CmpFront}.add$(1,\textit{min},i)$\COMMENT{Add dominating nature of $\textit{new}$, front index and solution index}
			\STATE break
		\ELSIF[$\textit{new}$ is dominated by $F_{\textit{min}}(i)$]{$\textit{isdom} = -1$}	
			\STATE \textit{CmpFront}.add$(-1,\textit{min},i)$\COMMENT{Add dominating nature of $\textit{new}$, front index and solution index}
			\STATE break
		\ELSE[$\textit{new}$ and $F_{\textit{min}}(i)$ are non-dominating]
			\STATE $\textit{count} \leftarrow \textit{count}{+}1$
		\ENDIF	
	\ENDFOR
	\IF{$\textit{count} = \vert F_{\textit{min}} \vert$}
		\STATE \textit{CmpFront}.add$(0,\textit{min},0)$\COMMENT{Add dominating nature of $\textit{new}$, front index and solution index}
	\ENDIF
\ELSE
	\STATE $\textit{mid} \leftarrow \lfloor \left( \textit{min} + \textit{max} \right) / 2 \rfloor$	  \COMMENT{Obtain the position of the node to be explored}
	\FOR{$i \leftarrow 1 \text{ to } \vert F_{\textit{mid}} \vert$}
		\STATE $\textit{isdom} \leftarrow${domNature$\left(\textit{new},F_{\textit{mid}}(i) \right)$}    \COMMENT{check the dominating nature of $\textit{new}$ and $F_{\textit{mid}}(i)$}		
		\IF[$\textit{new}$ dominates $\textit{sol}$]{$\textit{isdom} = 1$}		
			\STATE \textit{CmpFront}.add$(\textit{isdom},\textit{mid},i)$\COMMENT{Add dominating nature of $\textit{new}$, front index and solution index}
			\IF[Check for the existence of left sub-tree]{$\textit{mid} \neq \textit{min}$}
				\STATE $\text{InsertTree\_Right}(\mathcal{F},\textit{min},\textit{mid}{-}1,\textit{new})$\COMMENT{Explore left sub-tree}
			\ENDIF
			\STATE break
		\ELSIF[$\textit{sol}$ \textit{dominates} $\textit{new}$]{$F_{\textit{mid}}(i) = -1$}
			\STATE \textit{CmpFront}.add$(\textit{isdom},\textit{mid},i)$\COMMENT{Add dominating nature of $\textit{new}$, front index and solution index}
			\STATE $\text{InsertTree\_Right}(\mathcal{F},\textit{mid}{+}1,\textit{max},\textit{new})$\COMMENT{Explore right sub-tree}
			\STATE break
		\ELSE[$\textit{new}$ and $F_{\textit{mid}}(i)$ are non-dominating]
			\STATE $\textit{count} \leftarrow \textit{count}{+}1$
		\ENDIF	
	\ENDFOR
	\IF[All solutions in $F_{\textit{\textit{mid}}}$ are non-dominating with $\textit{new}$]{$\textit{count} = \vert F_{\textit{mid}} \vert$}
		\STATE \textit{CmpFront}.add$(0,\textit{mid},0)$\COMMENT{Add dominating nature of $\textit{new}$, front index and solution index}
		\IF{$\textit{mid} \neq \textit{min}$}
			\STATE $\text{InsertTree\_Right}(\mathcal{F},\textit{min},\textit{mid}{-}1,\textit{new})$\COMMENT{Explore left sub-tree}
		\ENDIF
	\ENDIF
\ENDIF
\end{algorithmic}
\caption{InsertTree\_Right$(\mathcal{F},\textit{min},\textit{max},\textit{new})$}
\label{alg:InsertTreeRight}
\end{algorithm*}

\subsection{Illustration of Algorithm \ref{alg:InsertTree}}
Algorithm \ref{alg:InsertTree} shows the insertion procedure of a new solution in the fronts using dominance binary search tree approach. Here like linear approach, we are also assuming that all the fronts are arranged in decreasing order of their dominance. This insertion can make the following changes in the tree.

\begin{enumerate}[I.]\compresslist
\item $\textit{new}$ can make new front i.e., creation of a new node.
\item $\textit{new}$ can be merged with any of the existing fronts.
\item $\textit{new}$ can be merged with any of the fronts by removing some of the solutions (which are dominated by $\textit{new}_{\textit{dom}}$) in that front. After this removal, the solutions in the lower dominance fronts are re-arranged.
\end{enumerate}

With the help of Algorithm \ref{alg:InsertTree} a new solution $\textit{new}$ is inserted in the given list of fronts. If there is only single front then Algorithm \ref{alg:InsertLinear} is called. Here the insertion is performed in the same manner as in linear approach. So in case of single front i.e. when all the solutions are non-dominating in nature then the linear approach and dominance tree based approach is same. If the number of fronts are more than one then the dominance binary search tree approach comes into picture. This algorithm first executes Algorithm \ref{alg:InsertTreeLeft} which gives a list known as \textit{CmpFront}. In place of Algorithm \ref{alg:InsertTreeLeft} which is based on Left Dominance Binary Search Tree, Algorithm \ref{alg:InsertTreeRight} can also be used which is based on Right Dominance Binary Search Tree.

Algorithm \ref{alg:InsertTreeLeft} and \ref{alg:InsertTreeRight} returns a list \textit{CmpFront}. Each element of this list is 3-folded - $<$dom,fIndex,sIndex$>$. `dom' is the dominating nature of $\textit{new}$ with a particular solution in the front. `fIndex' is the index of the front to which the $\textit{new}$ is compared and `sIndex' is the index of the solution in $F_{\text{fIndex}}$ to which $\textit{new}$ is compared and 'dom' is achieved. Thus `dom' shows the dominated nature of $\textit{new}$ with $F_{\text{fIndex}}(\text{sIndex})$. The `dom' can take three values $-1,0,1$. The value of `fIndex' can vary between $1$ to $K$. Generally the value of `sIndex' can vary between $1$ to $n_k, 1 \leq k \leq K$ but when $\textit{new}$ is non-dominating with all the solutions in a particular front then the value of `sIndex' is $0$. The maximum length of this list is $\lfloor \log_2K \rfloor + 1$ because the maximum number of fronts to which $\textit{new}$ can be compared is $\lfloor \log_2K \rfloor + 1$.

After obtaining \textit{CmpFront}, the correct position of $\textit{new}$ is to be identified. Let the length of \textit{CmpFront} is denoted by $\textit{len}$. There are following possibilities:

\noindent
$1)$ If last node in the \textit{CmpFront} is non-dominating with $\textit{new}$ then $\textit{new}$ will be inserted in the front corresponding to last node. The index of the front corresponding to last node is obtained by $\textit{CmpFront}[\textit{len}]_{\text{fIndex}}$.

\noindent
$2)$ If $\textit{new}$ dominates last node in the $\textit{CmpFront}$ then 
	\begin{enumerate}[I.]\compresslist
	\item Obtain the index of the front corresponding to last node. The index of this node, $\textit{fIndex} = \textit{CmpFront}[\textit{len}]_{\text{fIndex}}$.
	\item Obtain the index of the solution which is first dominated by $\textit{new}$ in the $F_{\textit{fIndex}}$. The index of this solution, $\textit{sIndex} = \textit{CmpFront}[\textit{len}]_{\text{sIndex}}$.
	\item Add the solution $F_{\textit{fIndex}}(\textit{sIndex})$ in $\textit{new}_{\textit{dom}}$
	\item Remove the solution $F_{\textit{fIndex}}(\textit{sIndex})$ from  $F_{\textit{fIndex}}$
	\item Obtain rest of the dominated solution in $F_{\textit{fIndex}}$ by $\textit{new}$ using \textit{DomSet()} procedure. The \textit{DomSet()} is called with following parameter - $F_{\textit{fIndex}}$, $\textit{new}$, $\textit{sIndex+1}$ and $\textit{new}_{\textit{dom}}$.
\item After obtaining $\textit{new}_{\textit{dom}}$, $\textit{new}$ is being added to $F_{\textit{fIndex}}$. Now one of the following three possibilities may arise:
	\begin{itemize}
	\item If $\textit{fIndex} = K$ then $\textit{new}_{\textit{dom}}$ creates a new front $F_{K{+}1}$ having lowest dominance.
	\item If $\textit{new}$ dominates all the solutions in front $F_{\textit{fIndex}}$ i.e. $\vert F_{\textit{fIndex}} \vert = 1$ then the dominance level of all the fronts $F_{\textit{fIndex}{+}1},F_{\textit{fIndex}{+}2},\ldots,F_K$ are increased by $1$ and $\textit{new}_{\textit{dom}}$ is assigned to front $F_{\textit{fIndex}{+}1}$. 
	\item If none of the above two conditions are satisfied then \textit{UpdateInsert()} procedure is used which re-arranges the solutions in their respective fronts.  The \textit{UpdateInsert()} is called with following parameter - $\mathcal{F}$,   $\textit{new}_{\textit{dom}}$ and $\textit{fIndex}{+}1$.
	\end{itemize}
	
	\end{enumerate}

\noindent
$3)$ If the values denoting dominance nature of $\textit{new}$ in \textit{CmpFront} list contains $-1$ for all the fronts i.e. $\textit{new}$ is dominated by all the compared fronts. Formally $\textit{CmpFront}[i]_{\text{dom}} = -1 \,\,\, \forall i,  1 \leq i \leq \textit{len}$. In this case it is clear that $\textit{new}$ will make another front $F_{K{+}1}$ which is having lower dominance than all the existing $K$ fronts. 

\noindent
$4)$ The \textit{CmpFront} list is traversed backward i.e. from the end to start. If the two consecutive values denoting dominance nature of $\textit{new}$ in  \textit{CmpFront} list are different then the position of $\textit{new}$ is identified otherwise we move to next \textit{CmpFront} element. Formally for the position of $\textit{new}$ is to be identified $\textit{CmpFront}[i]_{\text{dom}} \neq \textit{CmpFront}[\textit{i-}1]_{\text{dom}}$. If this condition is met and if the dominating value of next front (current front is i and next front is i-1) is $0$ then the $\textit{new}$ will be inserted in the next front. If dominating value of next front is $1$ then \textit{UpdateInsert()} procedure is called with following parameter - $\mathcal{F}$, $\textit{new}_{\textit{dom}}$ and $\textit{CmpFront}[\textit{i-1}]_{\text{fIndex}}$.

\subsection{Illustration of \textit{InsertTree\_Left}($\mathcal{F},\textit{min},\textit{max},\textit{new}$) procedure}
Initially $\textit{new}$ is compared with the root of the tree. The index of the root is calculated as $\textit{mid} = \lfloor (1+K)/2 \rfloor$. 

\noindent
1. If any solution in the root dominates $\textit{new}$ then the $\textit{new}$ will be having lower dominance than the root. So the algorithm explore the right sub-tree of the root using recursive procedure \textit{InsertTree\_Left}($\mathcal{F},\textit{mid}{+}1,\textit{max},\textit{new}$).

\noindent
2. If $\textit{new}$ and solution in the root is non-dominating then we continue dominance comparison of $\textit{new}$ with rest of the solutions in the root. If $\textit{new}$ is non-dominating with rest of the solutions in the root then  $\textit{new}$ can not have lower dominance than root. But it can have higher dominance than the root so for this the left sub-tree is explored using recursive procedure \textit{InsertTree\_Left}($\mathcal{F},\textit{min},\textit{mid}{-}1,\textit{new}$).

\noindent
3. If $\textit{new}$ dominates the solution in the node then the $\textit{new}$ has higher dominance than the root node so the left-sub tree of the root is explored using recursive procedure \textit{InsertTree\_Left}($\mathcal{F},1,\textit{mid}{-}1,\textit{new}$).

\noindent
\textbf{Terminating Condition of Algorithm \ref{alg:InsertTreeLeft} and \ref{alg:InsertTreeRight}:} The terminating condition depends on how the index of the root of a sub-tree i.e. $\textit{mid}$ is calculated. It depends on whether the tree is Left-Balanced or Right-Balanced. The procedure terminates when any one of the following conditions are satisfied.

\begin{enumerate}\compresslist
\item A leaf node is encountered.
\item If the tree is Left-Balanced Dominance Binary Search Tree and a node with single child is encountered which dominates $\textit{new}$.
\item If the tree is Right-Balanced Dominance Binary Search Tree and a node with single child is encountered which does not dominate $\textit{new}$.
\end{enumerate}

\noindent
\textbf{Complexity Analysis:} The maximum number of fronts to which $\textit{new}$ is compared is $\lfloor \log_2K \rfloor +1$. Each comparison to the front adds an element in the list $\textit{CmpFront}$. Thus the space complexity of the proposed algorithm is $\mathcal{O}(\log_2K)$. In the worst case, the $\textit{new}$ is to be checked for dominance with all the solutions in the compared fronts. If there is equal division of solutions in all the fronts i.e. each front has $\frac{N}{K}$ solutions then the complexity of the algorithm \ref{alg:InsertTreeLeft} and \ref{alg:InsertTreeRight} is $\mathcal{O}(M \times \frac{N}{K} \times \log K)$ which is $\mathcal{O}(MN)$ is worst case.

\subsection{Complexity Analysis using Dominance Binary Search Tree based Approach}
The \textit{InsertTree()} procedure first use either Algorithm \ref{alg:InsertTreeLeft} or \ref{alg:InsertTreeRight} which gives a list known as \textit{CmpFront}. The worst case time complexity of algorithm \ref{alg:InsertTreeLeft} is $\mathcal{O}(MN)$. After getting \textit{CmpFront}, this list is completely traversed at-most once. In this traversal either a new front is created or \textit{UpdateInsert()} procedure is called. Thus the worst case time complexity of Algorithm \ref{alg:InsertTree} is given by $\mathcal{O}(MN)$ + $\mathcal{O}(\log K)$ + $\mathcal{O}(MN^2)$ which is $\mathcal{O}(MN^2)$.  The space complexity of dominance tree based approach is $\mathcal{O}(\log K)$ which is used to store the list \textit{CmpFront}.

In case of left dominance binary search tree, the best case occurs when $\mathcal{F} = \left\lbrace F_1,F_2,F_3 \right\rbrace$ where $n_1=1, n_2=1, n_3=N-2$ and new solution $\textit{new}$ dominates the solution in both $F_1$ and $F_2$. In this case only two comparisons are required so the best case complexity is $\mathcal{O}(M)$.

In case of right dominance binary search tree, the best case occurs when $\mathcal{F} = \left\lbrace F_1,F_2 \right\rbrace$ where $n_1=1, n_2=N-1$ and new solution $\textit{new}$ dominates the solution in $F_1$. In this case only one comparison is  needed so the best case complexity is $\mathcal{O}(M)$.

\subsection{Comparison between Linear and Dominance Tree Based Approach}
The only difference between the linear and dominance tree based approach is the way to find the index of the front where \textit{UpdateInsert()} procedure is applied. For this purpose, the number of fronts to which $\textit{new}$ is compared depends on the dominance nature of $\textit{new}$ with all other solutions in case of linear approach. In case of dominance tree based approach, the number of fronts to which $\textit{new}$ is compared is either $\lfloor \log_2K \rfloor$ or $\lfloor \log_2K \rfloor + 1$. In case of full dominance binary search tree, the number of fronts to which $\textit{new}$ is compared is $\lfloor \log_2K \rfloor + 1$. When the tree is not fully balanced, the number of fronts is either $\lfloor \log_2K \rfloor$ or $\lfloor \log_2K \rfloor + 1$ depending on the $\textit{new}$.

\section{Look Up}\label{sec:lookup}
When a inferior solution is to be removed from $\mathcal{F}$ then first its location should be identified. So in this section we will discuss how to identify the location of a solution in $\mathcal{F}$. The linear approach proposed in \cite{li2014efficient}, checks for all the fronts in serial manner to identify a given solution. The number of comparison in various scenarios when this technique is followed is described next. 

\begin{enumerate}[I.]\compresslist
\item \textbf{All Solutions are dominating:} Here all the solutions are dominating in nature i.e. $N$ solutions are divided into $N$ fronts. The maximum number of comparison to search any solution is $N$. For this the searched solution is in the last front. The minimum comparison to search any solution is $1$. Here the searched solution is in the first front. 
\item \textbf{All Solutions are non-dominating:} Here all the solutions are non-dominating in nature i.e. all the $N$ solutions are in single front. The maximum number of comparison to search any solution is $N$. For this the searched solution is the last solution in the front. The minimum time to search any solution is $1$. Here the searched solution is the first solution in the front.
\item \textbf{Equal solutions in all the fronts:} When there are equal division of solutions in each front then the number of solutions in each front would be $\frac{N}{K}$. Assume each solution in a front is dominated by all solutions in the preceding front. Maximum number of comparison to search a solution is $K + \frac{N}{K}-1$. For this the searched solution is the last solution in last front. The minimum number of comparison to search any solution is $1$. Here the searched solution is the first solution in the first front.
\end{enumerate}

Two solutions are compared using CheckDom(A,B) procedure. The procedure is shown in Algorithm \ref{alg:CheckDom}. This procedure takes as input two solutions $A$ and $B$ and return the dominance relationship between them. When two solutions are compared then one of the following possibilities may arise:

\begin{algorithm}
\begin{algorithmic}[1]
\renewcommand{\algorithmicrequire}{\textbf{Input:}}
\REQUIRE $A$: First solution, $B$: Second Solution
\renewcommand{\algorithmicensure}{\textbf{Output:}}
\ENSURE Nature of $A$ with respect to $B$
\FOR{$i \text{ to } M$}
	\IF{$A_i < B_i$}
		\STATE $\textit{flag1} \leftarrow \textsc{True}$
	\ELSIF{$A_i > B_i$}
		\STATE $\textit{flag2} \leftarrow \textsc{True}$
	\ELSE
		\STATE $\textit{count++}$	
	\ENDIF
\ENDFOR
\IF{$\textit{flag1} = \textsc{True}$ \&\& $\textit{flag2} = \textsc{False}$}
	\RETURN $1$
\ELSE
	\IF{$\textit{flag1} = \textsc{False}$ \&\& $\textit{flag2} = \textsc{True}$}
		\RETURN $-1$
	\ELSE
		\IF{$\textit{count} = M$}
			\RETURN $2$
		\ELSE
			\RETURN $0$
		\ENDIF	
	\ENDIF
\ENDIF			
\end{algorithmic}
\caption{$\textit{CheckDom}(A,B)$}
\label{alg:CheckDom}
\end{algorithm}

\begin{itemize}\compresslist
\item Searched solution $\textit{sol}$ dominates the compared solution in the front. The function returns $1$. 
\item Searched solution $\textit{sol}$ is dominated by the compared solution in the front. The function returns $-1$. 
\item Searched solution $\textit{sol}$ is non-dominating with the compared solution in the front. The function returns $0$. 
\item Searched solution $\textit{sol}$ is same as the compared solution in the front. The function returns $2$. 
\end{itemize}

\noindent
\textbf{Complexity Analysis:} When two solutions are compared then at-most $M$ objectives are to be checked so the worst case complexity of this algorithm is $\mathcal{O}(M)$.

The proposed dominance tree based approach for locating a given solution in $\mathcal{F}$ is described in Algorithm \ref{alg:LookUp}. In this algorithm we have used the Left Dominance Binary Search Tree. The corresponding Right Dominance Binary Search Tree can also be used for lookup purpose. Here we first check the searched solution with the solutions in the root of the tree. The procedure $\textit{CheckDom()}$ is used to compare two solutions. When the searched solution is checked with solution in the root front then there are four possibilities: 

\begin{algorithm}
\begin{algorithmic}[1]
\renewcommand{\algorithmicrequire}{\textbf{Input:}}
\REQUIRE $\mathcal{F}$: Set of non-dominated fronts \\ 
\hspace{0.18 cm} $\textit{min}$: Lower index of the front \\
\hspace{0.13 cm} $\textit{max}$: Upper index of the front \\
\hspace{0.30 cm} $\textit{sol}$: Solution to be searched
\renewcommand{\algorithmicensure}{\textbf{Output:}}
\ENSURE Index of the searched solution 
\IF{$\textit{min} = \textit{max}$}
	\FOR{$i \leftarrow 1 \text{ to } \vert F_{\textit{min}} \vert$}
		\STATE $\textit{isDom} \leftarrow \textit{CheckDom}(sol,F_{\textit{min}}(i))$
		\IF{$\textit{isDom} = 1$}
		 	\RETURN $-1$
		 \ELSIF{$\textit{isDom} = -1$}
		 	\RETURN $-1$
		 \ELSIF{$\textit{isDom} = 2$}
		 	\RETURN $(\textit{min},i)$
		 \ENDIF
	\ENDFOR
\ELSE
	\STATE $\textit{mid} \leftarrow \lceil \frac{\textit{min}+\textit{max}}{2} \rceil$
	\FOR{$i \leftarrow 1 \text{ to } \vert F_{\textit{mid}} \vert$}
		 \STATE $\textit{isDom} \leftarrow \textit{CheckDom}(sol,F_{\textit{mid}}(i))$
		 \IF{$\textit{isDom} = 1$}
		 	\RETURN $\textit{LookUp}(\mathcal{F},\textit{min},\textit{mid}{-}1,\textit{sol})$
		 \ELSIF{$\textit{isDom} = -1$}
		 	\IF{$\textit{mid} \neq \textit{max}$}
		 		\RETURN $\textit{LookUp}(\mathcal{F},\textit{mid}{+}1,\textit{max},\textit{sol})$
		 	\ENDIF
		 \ELSIF{$\textit{isDom} = 2$}
		 	\RETURN $(\textit{mid},i)$
		 \ENDIF
	\ENDFOR 
	\RETURN $\textit{LookUp}(\mathcal{F},\textit{min},\textit{mid}{-}1,\textit{sol})$
\ENDIF
\RETURN $-1$
\end{algorithmic}
\caption{$\textit{LookUp}(\mathcal{F},\textit{min},\textit{max},\textit{sol})$}
\label{alg:LookUp}
\end{algorithm}

\begin{enumerate}\compresslist
\item If the searched solution dominates any solution in the root then searched solution is having higher dominance than the root so only left sub-tree of root is explored.
\item If the searched solution is dominated by any solution in the root then searched solution is having lower dominance than the root so only right sub-tree of root is explored.
\item If the searched solution is same as any solution in the root then the lookup process terminates with the index of the front and solution inside the front.
\item If searched solution is non-dominating with all the solutions in the root then searched solution can not have lower dominance than root so only left sub-tree of root is explored.
\end{enumerate}

Let the searched solution is same as $F_i(j)$ i.e. the searched solution is the $j$-th solution in $F_i$. The algorithm returns $(i,j)$. In case the solution is not present in $\mathcal{F}$ the algorithm returns $-1$. The number of comparison in various scenarios when dominance tree based technique is followed is described next.

\begin{enumerate}[I.]\compresslist
\item \textbf{All Solutions are dominating:} Here all the solutions are dominating in nature i.e. $N$ solutions are divided into $N$ fronts. The maximum number of comparison to search any solution is given by $\lfloor \log N \rfloor {+} 1$. For this the searched solution is in the leaf node at depth $\lfloor \log N \rfloor$ (In case of full dominance tree, the searched solution should be at any leaf node because all the leaf node have the same depth and i.e. $\lfloor \log N \rfloor$). The minimum time to search any solution is $1$. Here the searched solution is at the mid front. 

\item \textbf{All Solutions are non-dominating:} Here all the solutions are non-dominating in nature i.e. all the $N$ solutions are in single front. The maximum number of comparison to search any solution is given by $N$. For this the searched solution is the last solution in the front. The minimum time to search any solution is $1$. Here the searched solution is the first solution in the front.  

\item \textbf{Equal solutions in all the fronts:} When there are equal division of solutions in each front then the number of solutions in each front would be $\frac{N}{K}$. Assume each solution in a front is dominated by all solutions in the preceding front. Maximum number of comparison to search a solution is given by $\left( \lfloor \log K \rfloor {+} 1 \right) +  \frac{N}{K}-1 = \lfloor \log K \rfloor + \frac{N}{K}$. For this the searched solution should be the last solution in any leaf node which is at depth $\lfloor \log K \rfloor$ (In case of full dominance tree, for maximum number of comparison, the searched solution should be the last solution in any leaf node. This is because all the leaf node have the same depth and i.e. $\lfloor \log K \rfloor$). The minimum time to search any solution is $1$. Here the searched solution is the first solution in the first front.
\end{enumerate}

\subsection{Comparison}
When all the solutions are in the same front then both the approaches linear \cite{li2014efficient} as well as dominance tree based approach perform the same. But in case the number of fronts are $N$ then the number of comparison vary between $1$ to $N$ for linear approach while $1$ to $\lfloor \log N \rfloor {+} 1$ for dominance tree based approach. In this case in some situation linear approach performs better than the dominance tree based approach. Consider an example to illustrate such situation:

\begin{exmp}
Let there are $100$ solutions and these solutions are divided in $100$ fronts. In this case the number of comparison for dominance tree based approach vary between $1$ to $\lfloor \log 100 \rfloor + 1 = 7$. But in case of linear approach it vary between $1$ to $100$. Let we want to search the first solution, then the number of comparison using the linear approach is $1$ while using dominance tree based approach use $7$ comparison.
\end{exmp}

In general when all the solutions are dominating in nature and if the searched solution is among the first $\lfloor \log N \rfloor {+} 1$ solutions then the linear approach can outperform the dominance tree based approach otherwise the dominance tree based approach performs better. So when all the solutions are dominating in nature dominance tree based approach performs better for $N - \left( \lfloor \log N \rfloor {+} 1 \right)$ solutions and linear approach can perform better for $\lfloor \log N \rfloor {+} 1$ solutions.

%%%%%%%%%%%%%%
When there is equal division of elements in the fronts i.e. all the fronts have $\frac{N}{K}$ solutions and each solution in a front is dominated by all solutions in the preceding front. Then maximum number of comparison using linear approach would be $K+ \frac{N}{K}-1$ while maximum number of comparison using dominance tree based approach would be $\lfloor \log K \rfloor + \frac{N}{K}$. If the searched solution is among the first $\lfloor \log K \rfloor +1$ fronts then the linear approach can outperform the tree based approach otherwise the dominance tree based approach performs better. So in this case, dominance tree based approach performs better for $N - \frac{N}{K} \left( \lfloor \log K \rfloor {+} 1 \right)$ solutions and linear approach can perform better for $\frac{N}{K} \left( \lfloor \log K \rfloor {+} 1 \right)$ solutions.

\section{Delete a Solution}\label{sec:liDelete}
In this section we will discuss how the structure of the fronts changes after the removal of a solution $\textit{sol}$ from the set of fronts. Algorithm \ref{alg:Delete} shows the deletion procedure of a solution $\textit{sol}$ in the list of non-dominated fronts $\mathcal{F}$. This algorithm does not run the complete sorting algorithm again. It uses the non-dominance properties of the solution in the same front and uses the ranking of the front. Here we are assuming that all the fronts are arranged in decreasing order of their dominance.

First of all the position of the deleted solution is to be identified in $\mathcal{F}$. The position $\left( i,j \right)$ refers that the deleted solution is $j$-th solution in  $i$-th front. This operation is carried out using either linear search or \textit{LookUp()} procedure which is described in detail in Section \ref{sec:lookup}.

After the identification of the deleted solution, the solution is deleted from the front $F_i$. The removal of a solution from front $F_i$ requires re-arrangement of solutions in the fronts $F_i,F_{i+1}, F_{i+2}, \ldots, F_K$. If the deleted solution is from the last front i.e. $F_K$ then no solution is re-arranged and the process of deletion terminates. But if the deleted solution is in front $F_k, 1 \leq k < K$ then the re-arrangement of solutions occurs. This re-arrangement is performed by \textit{UpdateDelete()} procedure which is described in Algorithm \ref{alg:UpdateDelete}.

\begin{algorithm}
\begin{algorithmic}[1]
\renewcommand{\algorithmicrequire}{\textbf{Input:}}
\REQUIRE  $\mathcal{F} = \left\lbrace F_1,F_2,\ldots,F_K \right\rbrace$: Non-dominated fronts in the decreasing order of their dominance\\ 
\hspace{3.08 cm} $\textit{new}$: A new solution 
\renewcommand{\algorithmicensure}{\textbf{Output:}}
\ENSURE Updated $\mathcal{F}$ after removal of $\textit{sol}$
\STATE $(i,j) \leftarrow  \textit{LookUp}(\mathcal{F},\textit{sol})$
\STATE $F_i \leftarrow F_i \setminus F_i(j)$
\IF{$i \neq K$}
	\STATE $\textit{UpdateDelete}(\mathcal{F},i)$
\ENDIF
\end{algorithmic}
\caption{Delete$(\mathcal{F},\textit{sol})$}
\label{alg:Delete}
\end{algorithm}

\begin{algorithm}
\begin{algorithmic}[1]
\renewcommand{\algorithmicrequire}{\textbf{Input:}}
\REQUIRE $\mathcal{F}$: Set of non-dominated fronts \\ 
\hspace{0.0001 cm} $\textit{index}$: Non-domination level index
\renewcommand{\algorithmicensure}{\textbf{Output:}}
\ENSURE Updated $\mathcal{F}$ 
\STATE $l \leftarrow \vert F_\textit{index} \vert$
\FOR{$i \leftarrow 1 \text{ to } \vert F_{\textit{index}{+}1} \vert$}
	\STATE $\textit{count} \leftarrow 0$
	\FOR{$j \leftarrow 1 \text{ to } l$}
		\IF{$\textit{dominates}(F_{\textit{index}{+}1}(i), F_{\textit{index}}(j)) = 0$}
			\STATE $\textit{count} \leftarrow \textit{count} + 1$		
		\ENDIF
	\ENDFOR
	\IF{$\textit{count} = l$}
		\STATE $F_{\textit{index}} \leftarrow F_{\textit{index}} \cup F_{\textit{index}+1}(i)$
		\STATE $F_{\textit{index}+1} \leftarrow F_{\textit{index}+1} \setminus F_{\textit{index}+1}(i)$
		\STATE $i \leftarrow i - 1$
	\ENDIF
\ENDFOR
\IF{$F_{\textit{index}+1} = \Phi$}
	\STATE Decrease the domination level of front $F_k, k \in \left\lbrace \textit{index}{+}2, \textit{index}{+}3, \ldots, K \right\rbrace$	
\ELSIF{$\vert F_{\textit{index}} \vert = l$}
	\STATE // Do nothing process terminates
\ELSE
	\STATE $\textit{UpdateDelete}(\mathcal{F},\textit{index}{+}1)$
\ENDIF
\end{algorithmic}
\caption{UpdateDelete$(\mathcal{F},\textit{index})$}
\label{alg:UpdateDelete}
\end{algorithm}

\subsection{Illustration of UpdateDelete$(\mathcal{F},i)$ procedure}
This procedure takes as input the set of non-dominated fronts $\mathcal{F}$ and the index of the front $F_i$ from where the solution is deleted. This procedure updates $\mathcal{F}$ by either removing an existing front or by re-arranging the solutions within the existing fronts.

Initially the cardinality of $F_\textit{index}$ is stored in $l$. This is because when the solutions from $F_{\textit{index}{+}1}$ is compared with $F_\textit{index}$ for non-dominance then it should be compared with only first $l$ solutions. Here we find the solutions in $F_{\textit{index}+1}$ which are non-dominated with $F_\textit{index}$. The solutions which are non-dominated with $F_\textit{index}$ are added to it and removed from $F_{\textit{index}+1}$. This removal guarantees that no solution occupy more than one place. After this addition and removal, following cases can occur:

\begin{itemize}\compresslist
\item If all the solutions in front $F_{\textit{index}+1}$ are merged to front $F_\textit{index}$ i.e. $F_{\textit{index}+1} = \Phi$ then the non-domination level of  fronts $F_{\textit{index}+2},F_{\textit{index}+3},\ldots,F_{K}$ are decreased by $1$.
\item When no solution from front $F_{\textit{index}+1}$ is merged with $F_{\textit{index}}$ i.e. $\vert F_{\textit{index}} \vert = l$ then the process terminates.
\item If some of the solutions in front $F_{\textit{index}+1}$ are merged to front $F_\textit{index}$ i.e. $F_{\textit{index}+1} \neq \Phi$ then the procedure is repeated with UpdateDelete$(\mathcal{F},\textit{index}{+}1)$.
\end{itemize}

\noindent
\textbf{Complexity Analysis:}
In this algorithm the maximum number of comparison is performed when the deleted solution is from the front $F_1$ i.e. the procedure is called with $\textit{index}=1$. The maximum number of comparison occurs when each call to this procedure shifts one solution in higher level front after comparing with all the solutions. Thus the maximum number of comparison for this procedure is given by  $\left( n_1{-}1 \right)n_2+\left( n_2{-}1 \right)n_3+\ldots+\left( n_{k-1}{-}1 \right)n_K$. Each comparison between two solutions requires at-most $M$ comparison between $M$ objectives. Thus the worst case complexity of this procedure is $\mathcal{O}(MN^2)$.

\subsection{Complexity of Proposed Approach}
Here we will analyze the complexity of the proposed approach. We can see from all the algorithms \ref{alg:LookUp}, \ref{alg:Delete} and \ref{alg:UpdateDelete} that they involve scalar variables only except for the given set of fronts $F_k, 1 \leq k \leq K$, where $K$ is the number of fronts. Therefore, the space complexity of the proposed approach is $\mathcal{O}(1)$. The worst case time complexity of \textit{UpdateDelete()} procedure dominates the worst case time complexity of \textit{LookUp()} procedure. The complexity of \textit{UpdateDelete()} procedure is quadratic while \textit{LookUp()} has linear complexity. Thus the overall worst case complexity of the proposed approach is $\mathcal{O}(MN^2)$.

When sequential search strategy is used the the best case occurs when $F_1$ has single solution which is to be deleted. In this case only one comparison is required so the best case complexity is $\mathcal{O}(M)$.

When dominance tree based search strategy is used the the best case occurs when $F_{\textit{mid}}$ has single solution which is to be deleted. In this case only one comparison is required so the best case complexity is $\mathcal{O}(M)$.

\section{Number of Dominance Comparison}\label{sec:dominanceComparison}
In this section, we obtain the maximum number of dominance comparison occurred while inserting a new solution or deleting an existing solution in given set of fronts using the linear as well as dominance tree based approach. In case of deletion, the linear approach uses the sequential search for locating the solution while tree based approach uses the \textit{LookUp()} procedure for the same. 

\subsection{Linear Approach}
Here we obtain the maximum number of dominance comparison when either a new solution $\textit{new}$ is being inserted in $\mathcal{F}$ or an existing solution $\textit{sol}$ is being deleted from $\mathcal{F}$ using linear approach.

\subsubsection{Insert}
For maximum number of dominance comparison in linear approach, the new solution $\textit{new}$ dominates $n_1{-}1$ solutions in the first front. So maximum number of dominance comparison is given by Equation \ref{eq:linearMaximum}. In this case the \textit{UpdateInsert()} procedure is called with index value $2$.
\begin{multline}\label{eq:linearMaximum}
\#\textit{Comp}_{\textit{Linear}} = n_1 + \left[ \left( n_1{-}1 \right).n_2 + \ldots + \left( n_{K{-}1} {-} 1 \right).n_K  \right]
\end{multline}

$\#\textit{Comp}_{\textit{Linear}}$ attains its maximum value when there are exactly two fronts. For proof see Appendix \ref{appendix:linear}. In case of even number of solutions, first front should have $\frac{N}{2}{+}1$ solutions while second front should have $\frac{N}{2}{-}1$ solutions. In case of odd number of solutions, the first front has $\lceil\frac{N}{2}\rceil$ solutions and second front has $\lfloor\frac{N}{2}\rfloor$. In this way the maximum number of comparisons is 

\noindent
\textbf{N is Even:} 

$\#\textit{Comp}_{\textit{Linear}} = \left( \frac{N}{2}{+}1 \right) + \left[ \left( \frac{N}{2} {+} 1 \right) {-} 1 \right] \left( \frac{N}{2} {-} 1 \right) = \frac{N^2}{4}{+}1$

\noindent
\textbf{N is Odd:} 

$\#\textit{Comp}_{\textit{Linear}} = \lceil \frac{N}{2} \rceil +  \left[ \lceil \frac{N}{2} \rceil-1 \right] \lfloor \frac{N}{2} \rfloor $

\hspace{1.92 cm}$= \frac{N+1}{2} + \left[ \frac{N+1}{2} -1 \right] \frac{N-1}{2} = \frac{N^2+3}{4} = \lceil \frac{N^2}{4} \rceil$

\[ \#\textit{Comp}_{\textit{Linear}} =
  \begin{cases}
    \frac{N^2}{4}{+}1       & \quad \text{if } N \text{ is even}\\
     \lceil \frac{N^2}{4} \rceil    & \quad \text{if } N \text{ is odd}\\
  \end{cases}
\]

\subsubsection{Delete}
For maximum number of dominance comparison the value of Equation \ref{eq:linearMaximum} should be maximized. Here the deleted solution $sol$ is the last solution in the first front. So the \textit{UpdateDelete()} procedure is called with index value $1$.

\subsection{Left Dominance Binary Search Tree Based Approach}
Here the maximum number of dominance comparison in case of insertion and deletion of a solution is obtained when Left Dominance Binary Search Tree based approach is used.

\subsubsection{Insert}
The index of the root front is obtained by $\textit{mid} = \lceil \frac{1+N}{2} \rceil$. The height of the dominance tree $h = \lfloor \log N \rfloor$. For maximum number of dominance comparison, the new solution $\textit{new}$ dominates the $n_1-1$ solutions in the first front. So, maximum number of dominance comparison is given by Equation \ref{eq:leftTreeMaximum}.
\begin{multline}\label{eq:leftTreeMaximum}
\#\textit{Comp}_{\textit{LeftTree}}  = \left[ n_{\lceil \frac{\textit{mid}}{2^0} \rceil} + n_{\lceil \frac{\textit{mid}}{2^1} \rceil} + n_{\lceil \frac{\textit{mid}}{2^2} \rceil} + \ldots + n_{{\lceil \frac{\textit{mid}}{2^{h}} \rceil}} \right] + \\ 
\left[ \left( n_1{-}1 \right).n_2 + \left( n_2{-}1 \right).n_3 + \ldots + \left( n_{K{-}1}{-}1 \right).n_K  \right]
\end{multline}

This value will be maximum when there are exactly two fronts. For proof see Appendix \ref{appendix:left}. In case of even number of solutions, first front should have $\frac{N}{2}+1$ solutions while second front should have $\frac{N}{2}-1$ solutions. In case of odd number of solutions, the first front has $\lceil \frac{N}{2} \rceil$ solutions and second front has $\lfloor \frac{N}{2} \rfloor$. In this way the maximum number of comparison is 

\noindent
\textbf{N is Even:} 

%\noindent
$\#\textit{Comp}_{\textit{LeftTree}} = \left( \frac{N}{2}{-}1 \right) {+} \left( \frac{N}{2}{+}1 \right) {+} \left[ \left( \frac{N}{2} {+}1 {-} 1 \right) \left( \frac{N}{2}{-}1 \right) \right]$ %= \frac{N^2}{4} + \frac{N}{2}$

\hspace{2.1 cm}$=\frac{N^2}{4} + \frac{N}{2}$

\noindent
\textbf{N is Odd:} 

$\#\textit{Comp}_{\textit{LeftTree}} = \lfloor \frac{N}{2} \rfloor +   \lceil \frac{N}{2} \rceil + \left[ \left( \lceil \frac{N}{2} \rceil-1 \right) \lfloor \frac{N}{2} \rfloor \right]$

\hspace{2.05 cm}$= N + \left[ \left( \frac{N+1}{2} -1 \right) \frac{N-1}{2} \right]$

\hspace{2.05 cm}$= \frac{N^2}{4} + \frac{N}{2} + \frac{1}{4}$

\subsubsection{Delete}
For maximum number of dominance comparison, the deleted solution $\textit{sol}$ has to be the last solution in first front $F_1$. So, maximum number of dominance comparison is given by Equation \ref{eq:leftTreeMaximum}. Here the deleted solution is located using left dominance binary search tree.

\subsection{Right Dominance Binary Search Tree Based Approach}
Here the maximum number of dominance comparison in case of insertion and deletion of a solution is obtained when Right Dominance Binary Search Tree based approach is used.

\subsubsection{Insert}
The index of the root front is obtained by $\textit{mid} = \lfloor \frac{1+N}{2} \rfloor$.  The height of the dominance tree $h = \lfloor \log N \rfloor$. For maximum number of dominance comparison, the new solution $\textit{new}$ dominates the solutions in the first front. So, maximum number of dominance comparison is given by Equation \ref{eq:rightTreeMaximum}. 
\begin{multline}\label{eq:rightTreeMaximum}
\#\textit{Comp}_{\textit{RightTree}}  = \left[ n_{\lfloor \frac{\textit{mid}}{2^0} \rfloor} + n_{\lfloor \frac{\textit{mid}}{2^1} \rfloor} + \ldots + n_{{\lfloor \frac{\textit{mid}}{2^{h}} \rfloor}} \right] + \\ 
\left[ \left( n_1-1 \right).n_2 + \left( n_2-1 \right).n_3 + \ldots + \left( n_{K-1}-1 \right).n_K  \right]
\end{multline}

This value will be maximum when there are exactly two fronts. For proof see Appendix \ref{appendix:right}. In case of even number of solutions, first front should have $\frac{N}{2}+1$ solutions while second front should have $\frac{N}{2}-1$ solutions. In case of odd number of solutions, the first front has $\lceil \frac{N}{2} \rceil$ solutions and second front has $\lfloor \frac{N}{2} \rfloor$. In this way the maximum number of comparison is

\noindent
\textbf{N is Even:} 

$\#\textit{Comp}_{\textit{RightTree}} = \left( \frac{N}{2} + 1 \right) + \left[ \left( \frac{N}{2} + 1 - 1 \right) \left( \frac{N}{2} - 1 \right) \right]$

\hspace{2.22 cm}$= \frac{N^2}{4}+1$

\noindent
\textbf{N is Odd:} 

$\#\textit{Comp}_{\textit{LeftTree}} = \lceil \frac{N}{2} \rceil + \left[ \left( \lceil \frac{N}{2} \rceil-1 \right) \lfloor \frac{N}{2} \rfloor \right]$

\hspace{2.05 cm}$= \frac{N+1}{2} + \left[ \left( \frac{N+1}{2} -1 \right) \frac{N-1}{2} \right]$

\hspace{2.05 cm}$= \frac{N^2+3}{4} = \lceil \frac{N^2}{4} \rceil$

\subsubsection{Delete}
For maximum number of dominance comparison, the deleted solution $\textit{sol}$ has to be the last solution in the first front $F_1$. So, maximum number of dominance comparison is given by Equation \ref{eq:rightTreeMaximum}. Here the deleted solution is located using right dominance binary search tree.

\section{Case Study: All Solutions are non-dominated}\label{sec:nd}
In this section, we will discuss the maximum and minimum number of dominance comparison needed when either a new solution is inserted or an existing solution is deleted from the set of fronts where there is single front containing all the $N$ solutions. In this case the linear as well as dominance tree based approach performs the same. 

\subsection{Insert}
The maximum and minimum number of dominance comparison is discussed here in case of insertion. The value of $K=1$. Table \ref{tab:ndimax} shows this scenario. The changed structure in the front after insertion is also shown.

\begin{table}[h]
	\begin{center}
    \begin{subtable}[h]{0.15\textwidth}
        \centering
        \scalebox{0.70}{
        \begin{tabular}{| c | c |}
       		\hline
         	$\mathbfcal{F}$    &  \textbf{Solutions} \\ \hline
			$F_1$    &  $sol_1,sol_2,\ldots,sol_N$  \\ \hline
        \end{tabular}}
        \caption{}
        \label{tab:ndimax}
    \end{subtable}
    ~~
    \begin{subtable}[h]{0.15\textwidth}
        \centering
        \scalebox{0.70}{
        \begin{tabular}{| c | c |}
        	\hline
         	$\mathbfcal{F}$    &  \textbf{Solutions}    \\ \hline
        	$F_1$    &  $sol_1,sol_2,\ldots,sol_N,\textit{new}$ \\ \hline
        \end{tabular}}
        \caption{}
        \label{tab:ndimaxa}
    \end{subtable}
    ~~
    \begin{subtable}[h]{0.15\textwidth}
        \centering
        \scalebox{0.70}{
        \begin{tabular}{| c | c |}
        \hline
        $\mathbfcal{F}$    &  \textbf{Solutions} \\ \hline
		$F_1$    &  $\textit{sol}_1,\textit{sol}_2,\ldots,\textit{sol}_N$  \\  
		$F_2$    &  $\textit{new}$                       \\ \hline
        \end{tabular}}
        \caption{}
        \label{tab:ndimaxb}
    \end{subtable}
    \end{center}
    \begin{center}
    \begin{subtable}[h]{0.15\textwidth}
        \centering
        \scalebox{0.70}{
        \begin{tabular}{| c | c |}
			\hline
			$\mathbfcal{F}$    &  \textbf{Solutions}        \\  \hline
			$F_1$    &  $\textit{sol}_1,\textit{sol}_2,\ldots,\textit{sol}_p,\textit{new}$     \\ 
			$F_2$    &  $\textit{sol}_{p+1},\textit{sol}_{p+2},\ldots,\textit{sol}_N$ \\  \hline
        \end{tabular}}
        \caption{}
        \label{tab:ndimaxc}
    \end{subtable}
    ~~
    \begin{subtable}[h]{0.15\textwidth}
        \centering
        \scalebox{0.70}{
        \begin{tabular}{| c | c |}
        \hline
        $\mathbfcal{F}$    &  \textbf{Solutions} \\  \hline
		$F_1$    &  $\textit{new}$ \\
		$F_2$    &  $\textit{sol}_1,\textit{sol}_2,\ldots,\textit{sol}_N$  \\  \hline
        \end{tabular}}
        \caption{}
        \label{tab:ndimaxd}
    \end{subtable}
    ~~
    \begin{subtable}[h]{0.15\textwidth}
        \centering
        \scalebox{0.70}{
        \begin{tabular}{| c | c |}
       		\hline
			$\mathbfcal{F}$    &  \textbf{Solutions} \\ \hline
			$F_1$    &  $\textit{sol}_1,\textit{sol}_2,\ldots,\textit{sol}_N$  \\  
			$F_2$    &  $\textit{new}$                       \\ \hline
        \end{tabular}}
        \caption{}
        \label{tab:ndimine}
    \end{subtable}
    \end{center}   
    \begin{center}
    \begin{subtable}[h]{0.15\textwidth}
        \centering
        \scalebox{0.70}{
        \begin{tabular}{| c | c |}
        	\hline
         	$\mathbfcal{F}$    &  \textbf{Solutions}    \\ \hline
        	$F_1$    &  $sol_1,sol_2,\ldots,sol_{N{-}1}$ \\ \hline
        \end{tabular}}
        \caption{}
        \label{tab:nddmaxa}
    \end{subtable}
    ~~
    \begin{subtable}[h]{0.15\textwidth}
        \centering
        \scalebox{0.70}{
        \begin{tabular}{| c | c |}
        	\hline
       		$\mathbfcal{F}$    &  \textbf{Solutions} \\  \hline
			$F_1$    &  $sol_2,\ldots,sol_N$ \\ \hline
        \end{tabular}}
        \caption{}
        \label{tab:nddminb}
    \end{subtable} 
	\end{center}     
%    \caption{(a). All the solutions are non-dominating. (b). $\textit{new}$ is non-dominating with each solution in the front. (c). $\textit{new}$ is non-dominating with first $N{-}1$ solutions and is dominated by $N$-th solution. (d). $\textit{new}$ is non-dominated with first $p$ solutions and it dominates rest $N{-}p$ solutions. (e). $\textit{new}$ dominates all the solutions in the front. (f). $\textit{new}$ is dominated by first solution in the front. (g). $sol_N$ is deleted from the front. (h). $sol_1$ is deleted from the front.}
\caption{Initial and changed structure of fronts using linear approach when all the solutions are non-dominating.}
\end{table}

\noindent
\textbf{Maximum Number of Dominance Comparison:} 
The maximum number of dominance comparison is $N$. Here $\textit{new}$ is compared with all the solutions in the front. In this case there are four possibilities. 

\begin{enumerate}[I.]\compresslist
\item $\textit{new}$ will be merged to the front if it is non-dominating with each solution in the front. This is shown in Table \ref{tab:ndimaxa}.
\item $\textit{new}$ makes another front having lower dominance (lower rank) than current front if it is non-dominating with first $N-1$ solutions and is dominated by the $N$-th solution. See Table \ref{tab:ndimaxb} for this.
\item $\textit{new}$ merges in the same front and the solutions which are dominated by $\textit{new}$ make another front having lower dominance than current front. This is shown in Table \ref{tab:ndimaxc}.
\item $\textit{new}$ makes another front having higher dominance than the current front. This is possible when $\textit{new}$ dominates all the solutions in the front. See Table \ref{tab:ndimaxd}.
\end{enumerate}

\noindent
\textbf{Minimum Number of Dominance Comparison:} 
The minimum number of dominance comparison is $1$. Here $\textit{new}$ is compared with only first solution in the front. $\textit{new}$ makes another front having lower dominance than the current front because  $\textit{new}$ is dominated by first solution in the front. Refer Table \ref{tab:ndimine}.

\subsection{Delete}
The maximum and minimum number of dominance comparison is discussed here in case of deletion of an existing solution. The changed structure in the front after insertion is also shown.

\noindent
\textbf{Maximum Number of Dominance Comparison:} 
Maximum number of comparison occurs when the last solution i.e. $\textit{sol}_N$ is deleted from the front. Table \ref{tab:nddmaxa} shows this scenario.

\noindent
\textbf{Minimum Number of Dominance Comparison:} 
Minimum number of comparison occurs when the first solution i.e. $\textit{sol}_1$ is deleted from the front. Table \ref{tab:nddminb} shows this scenario.

\section{Case Study: All Solutions are dominated}\label{sec:d}
In this section, we will discuss the maximum and minimum number of dominance comparison needed when either a new solution is inserted or an existing solution is deleted from the set of fronts where there are N fronts. Each front contains single solution. 

\subsection{Linear}
The number of comparison using linear approach is discussed.

\subsubsection{Insert}
The maximum and minimum number of dominance comparison is discussed here in case of insertion. The value of $K=N$. Table \ref{tab:dimax} shows this scenario. The changed structure in the front after insertion is also shown.

\noindent
\textbf{Maximum Number of Dominance Comparison:} 
The maximum number of dominance comparison is $N$. Here $\textit{new}$ will be compared with solution in each front. In this case there are two possibilities. 

\begin{enumerate}[I.]\compresslist
\item $\textit{new}$ will be merged to the last front if $\textit{new}$ is dominated by the solutions in the first $N-1$ fronts and it is non-dominating with the solution in the last front. See Table \ref{tab:dimaxa}.
\item $\textit{new}$ makes another front if $\textit{new}$ is dominated by the solutions in all the $N$ fronts. The dominance of $\textit{new}$ will be the lowest among all the fronts. Refer Table \ref{tab:dimaxb}.
\end{enumerate}

\noindent
\textbf{Minimum Number of Dominance Comparison:} 
When all the solutions belong to different front then the minimum number of dominance comparison will be $1$. Here $\textit{new}$ will be compared with only single solution. In this case there are two possibilities.

\begin{enumerate}[I.]\compresslist
\item $\textit{new}$ will be merged to the first front if $\textit{new}$ is non-dominating with the solution in the first front. Refer Table \ref{tab:diminc}.
\item $\textit{new}$ makes another front (having higher dominance than the first front) if it dominates the solution in the first front. See Table \ref{tab:dimind}.
\end{enumerate}

\begin{table}[h]
	\begin{center}
    	\begin{subtable}[h]{0.10\textwidth}
        \centering
        \scalebox{0.75}{
        \begin{tabular}{| c | c |}
       		\hline
         	\textbf{Front}    &  \textbf{Solutions} \\  \hline
			$F_1$    &  $sol_1$   \\ 
			$F_2$    &  $sol_2$   \\ 
			\vdots   &   \vdots   \\ 
			$F_N$    &  $sol_N$   \\  \hline
        \end{tabular}}
        \caption{}
        \label{tab:dimax}
    	\end{subtable}
    	~~
    	\begin{subtable}[h]{0.10\textwidth}
        \centering
        \scalebox{0.75}{
        \begin{tabular}{| c | c |}
        	\hline
         	\textbf{Front}    &  \textbf{Solutions} \\  \hline
			$F_1$    &  $sol_1$     \\ 
			$F_2$    &  $sol_2$     \\ 
			\vdots   &  \vdots     \\ 
			$F_N$    &  $sol_N,\textit{new}$ \\  \hline
        \end{tabular}}
        \caption{}
        \label{tab:dimaxa}
    	\end{subtable}
    	~~
    	\begin{subtable}[h]{0.10\textwidth}
        \centering
        \scalebox{0.75}{
        \begin{tabular}{| c | c |}
       		\hline
         	\textbf{Front}    &  \textbf{Solutions} \\  \hline
			$F_1$      &  $sol_1$      \\ 
			$F_2$      &  $sol_2$      \\  
			\vdots     &   \vdots      \\ 
			$F_N$      &  $sol_N$      \\ 
			$F_{N+1}$  &  $\textit{new}$        \\   \hline
        \end{tabular}}
        \caption{}
        \label{tab:dimaxb}
    	\end{subtable}
    	~~
    	\begin{subtable}[h]{0.10\textwidth}
        \centering
        \scalebox{0.75}{
        \begin{tabular}{| c | c |}
       		\hline
         	\textbf{Front}    &  \textbf{Solutions} \\  \hline
			$F_1$      &  $sol_1,\textit{new}$      \\ 
			$F_2$      &  $sol_2$      \\  
			\vdots     &   \vdots      \\ 
			$F_N$      &  $sol_N$      \\ \hline
        \end{tabular}}
        \caption{}
        \label{tab:diminc}
    	\end{subtable}
    \end{center}
    \begin{center}
    	\begin{subtable}[h]{0.10\textwidth}
        \centering
        \scalebox{0.75}{
        \begin{tabular}{| c | c |}
        	\hline
         	\textbf{Front}    &  \textbf{Solutions} \\  \hline
         	$F_1$      &  $\textit{new}$        \\  
			$F_2$      &  $sol_1$      \\ 
			$F_3$      &  $sol_2$      \\  
			\vdots     &   \vdots      \\ 
			$F_{N+1}$  &  $sol_N$      \\ \hline
        \end{tabular}}
        \caption{}
        \label{tab:dimind}
    	\end{subtable} 
		~~
    	\begin{subtable}[h]{0.10\textwidth}
        \centering
        \scalebox{0.75}{
        \begin{tabular}{| c | c |}
        	\hline
         	\textbf{Front}    &  \textbf{Solutions} \\  \hline
			$F_1$    &  $sol_1$     \\ 
			$F_2$    &  $sol_2$     \\ 
			\vdots   &  \vdots     \\ 
			$F_{N{-}1}$    &  $sol_{N{-}1}$ \\  \hline
        \end{tabular}}
        \caption{}
        \label{tab:ddmaxa}
    	\end{subtable}
    	~~
    	\begin{subtable}[h]{0.10\textwidth}
        \centering
        \scalebox{0.75}{
        \begin{tabular}{| c | c |}
       		\hline
         	\textbf{Front}    &  \textbf{Solutions} \\  \hline
			$F_1$      &  $sol_2$      \\ 
			$F_2$      &  $sol_3$      \\  
			\vdots     &   \vdots      \\ 
			$F_{N-1}$      &  $sol_N$      \\ \hline
        \end{tabular}}
        \caption{}
        \label{tab:ddminb}
    	\end{subtable}
    \end{center}
%    \caption{(a). All the solutions are dominating. (b). $\textit{new}$ dominates the solutions in first $N-1$ fronts and it is non-dominating with solution in the last front. (c). $\textit{new}$ dominates the solutions in all $N$ fronts. (d). $\textit{new}$ is non-dominated with the solution in the first front. (e). $\textit{new}$ dominates the solution in the first front. (f). $sol_N$ is deleted from the front. (g). $sol_1$ is deleted from the front.}
\caption{Initial and changed structure of fronts using linear approach when all the solutions are dominating.}
\end{table}

\subsubsection{Delete}
The maximum and minimum number of dominance comparison is discussed here in case of deletion of an existing solution. The changed structure in the front after insertion is also shown.

\noindent
\textbf{Maximum Number of Dominance Comparison:} 
The maximum number of dominance comparison will be $N$. This occurs when the deleted solution is in the last front i.e. $sol_N$ is being deleted from the front. Table \ref{tab:ddmaxa} shows this scenario.

\noindent
\textbf{Minimum Number of Dominance Comparison:} 
The minimum number of dominance comparison will be $1$ and it occurs when the deleted solution is in the first front i.e. $sol_1$ is being deleted from the front. Table \ref{tab:ddminb} shows this scenario.

\subsection{Dominance Tree Based Approach}
The number of comparison using dominance tree based approach is discussed.

\subsubsection{Insert}
The maximum and minimum number of dominance comparison is discussed here in case of insertion. The value of $K=N$. Table \ref{tab:tdi} shows this scenario. The changed structure in the front after insertion is also shown. In this case the number of dominance comparison will be $\lfloor \log_2 N \rfloor + 1$ in all the cases. Here $\textit{new}$ will be compared with solution in $\lfloor \log_2 N \rfloor + 1$ front (there is only one solution in each front). In this case there are two possibilities.

\begin{table}[t]
    \begin{subtable}[h]{0.15\textwidth}
        \centering
        \scalebox{0.75}{
        \begin{tabular}{| c | c |}
       		\hline
         	\textbf{Front}    &  \textbf{Solutions} \\  \hline
			$F_1$    &  $sol_1$   \\ 
			$F_2$    &  $sol_2$   \\ 
			\vdots   &  \vdots    \\ 
			$F_N$    &  $sol_N$   \\  \hline
        \end{tabular}}
        \caption{}
        \label{tab:tdi}
    \end{subtable}
    \hfill
    \begin{subtable}[h]{0.15\textwidth}
        \centering
        \scalebox{0.75}{
        \begin{tabular}{| c | c |}
        	\hline
         	\textbf{Front}    &  \textbf{Solutions} \\  \hline
			$F_1$    &  $sol_1$   \\ 
			$F_2$    &  $sol_2$   \\ 
			$\dot{.}$  & $\dot{.}$ \\
			$F_p$    &  $sol_p,\textit{new}$   \\
			$\dot{.}$  & $\dot{.}$ \\
			$F_N$    &  $sol_N$   \\  \hline
        \end{tabular}}
        \caption{}
        \label{tab:tdia}
    \end{subtable}
    \hfill
    \begin{subtable}[h]{0.15\textwidth}
        \centering
        \scalebox{0.75}{
        \begin{tabular}{| c | c |}
       		\hline
         	\textbf{Front}    &  \textbf{Solutions} \\  \hline
			$F_1$      &  $sol_1$      \\ 
			$F_2$      &  $sol_2$      \\  
			$\dot{.}$  & $\dot{.}$ \\
			$F_N$      &  $sol_N$      \\ 
			$F_{N+1}$  &  $\textit{new}$        \\   \hline
        \end{tabular}}
        \caption{}
        \label{tab:tdib}
    \end{subtable}
    \\
    \begin{center}
    \begin{subtable}[h]{0.15\textwidth}
        \centering
        \scalebox{0.75}{
        \begin{tabular}{| c | c |}
       		\hline
         	\textbf{Front}    &  \textbf{Solutions} \\  \hline
         	$F_1$      &  $\textit{new}$        \\
			$F_2$      &  $sol_1$      \\ 
			$F_3$      &  $sol_2$      \\  
			\vdots     &   \vdots      \\ 
			$F_{N+1}$      &  $sol_N$      \\ \hline
        \end{tabular}}
        \caption{}
        \label{tab:tdic}
    \end{subtable}
     \hspace{0.25 cm}
    \begin{subtable}[h]{0.15\textwidth}
        \centering
        \scalebox{0.75}{
        \begin{tabular}{| c | c |}
        	\hline
         	\textbf{Front}    &  \textbf{Solutions} \\  \hline
			$F_1$      &  $sol_1$      \\
			$\dot{.}$  & $\dot{.}$    \\     
			$F_p$      &  $sol_p$      \\  			
			$F_{p+1}$  &  $\textit{new}$        \\  
			$\dot{.}$  & $\dot{.}$ \\    
			$F_{N+1}$  &  $sol_N$      \\ \hline
        \end{tabular}}
        \caption{}
        \label{tab:tdie}
    \end{subtable} 
    \end{center}
%    \caption{(a). All the solutions are dominating. (b). $\textit{new}$ is dominated by the solutions in first $p-1$ fronts and it is non-dominating with the solution in the $p$-th front. (c). $\textit{new}$ is dominated by the solutions in all $N$ fronts. (d). $\textit{new}$ dominates the solution in the first front. (e). $\textit{new}$ is dominated by the solution in first $p$ front and it dominates the solution in the $(p+1)$-th front.}
\caption{Initial and changed structure of fronts using linear approach when all the solutions are dominating.}
\end{table}

\begin{enumerate}[I.]\compresslist
\item $\textit{new}$ will be merged to any front. See Table \ref{tab:tdia}.
\item $\textit{new}$ makes another front.
	\begin{enumerate}[i.]\compresslist
	\item This new front can have lower dominance than the last front. See Table \ref{tab:tdib}.
	\item This new front can have higher dominance than the first front. See Table \ref{tab:tdic}.
	\item This new front can have dominance in between the first and the last front. See Table \ref{tab:tdie}.
	\end{enumerate}
\end{enumerate}

\subsubsection{Delete}
The maximum and minimum number of dominance comparison is discussed here in case of deletion. The value of $K=N$. 

\noindent
\textbf{Maximum Number of Dominance Comparison:}  
The maximum number of dominance comparison will be $\lfloor \log_2 N \rfloor + 1$. Maximum number of comparison occurs when the deleted solution is at the leaf of the tree.
 
\noindent
\textbf{Minimum Number of Dominance Comparison:} 
The minimum number of dominance comparison will be $1$. Minimum number of comparison occurs when the deleted solution is at the root of the tree.

\begin{table*}
\centering
\begin{tabular}{|c|c|c|c|c|c|}
\hline
\multirow{2}{*}{\textbf{Approach}} & \multirow{2}{*}{\textbf{Space Complexity: Worst Case}} & \multicolumn{2}{c|}{\textbf{Time Complexity}} & \multicolumn{2}{c|}{\textbf{Maximum Number of comparison}} \\ \cline{3-6}
                          &                                   & \textbf{Best Case}  & \textbf{Worst Case}  & \textbf{Insert} & \textbf{Delete} \\ \hline

Naive approach &  $\mathcal{O}(N)^*$ & $\mathcal{O}(MN^2)$ & $\mathcal{O}(MN^3)$ & $\frac{N(N+1)(N+2)}{6}$ & $\frac{(N-2)(N-1)N}{6}$ \\ \hline

Fast Non-dominated Sort & $\mathcal{O}(N^2)^*$ & $\mathcal{O}(MN^2)$ & $\mathcal{O}(MN^2)$ & $(N+1)N$ & $(N-1)(N-2)$\\ \hline

Climbing Sort & $\mathcal{O}(N^2)^*$ &  & $\mathcal{O}(MN^2)$ & $(N+1)N$ & $(N-1)(N-2)$\\ \hline

Deductive Sort & $\mathcal{O}(N)^*$ & $\mathcal{O}(MN\sqrt{N})$ & $\mathcal{O}(MN^2)$ & $\frac{(N+1)N)}{2}$ & $\frac{(N-1)(N-2)}{2}$  \\ \hline

Arena's Sort & $\mathcal{O}(N)^*$ & $\mathcal{O}(MN\sqrt{N})$ & $\mathcal{O}(MN^2)$ & $\frac{(N+1)N}{2}$ & $\frac{(N-1)(N-2)}{2}$  \\ \hline

ENS-SS & $\mathcal{O}(1)^*$ & $\mathcal{O}(MN\sqrt{N})$ & $\mathcal{O}(MN^2)$ & $\frac{(N+1)N}{2}$ & $\frac{(N-1)(N-2)}{2}$  \\ \hline

ENS-BS & $\mathcal{O}(1)^*$ & $\mathcal{O}(MN\log N)$ & $\mathcal{O}(MN^2)$ & $\frac{(N+1)N}{2}$ & $\frac{(N-1)(N-2)}{2}$  \\ \hline

%ENLU & $\mathcal{O}(N)^*$ & $\mathcal{O}(M)$  & $\mathcal{O}(MN^2)$ &  \\ \hline

\multirow{2}{*}{ENLU} & \multirow{2}{*}{$\mathcal{O}(N)^*$} & \multirow{2}{*}{$\mathcal{O}(M)$}  & \multirow{2}{*}{$\mathcal{O}(MN^2)$} & N Even: $\frac{N^2}{4} {+} 1$ & N Even: $\frac{N^2}{4} {+} 1$ \\ \cline{5-6}
                        &         & & &  N Odd: $\lceil \frac{N^2}{4} \rceil$ & N Odd: $\lceil \frac{N^2}{4} \rceil$ \\ \hline

\multirow{2}{*}{Linear} & \multirow{2}{*}{$\mathcal{O}(1)^*$} & \multirow{2}{*}{$\mathcal{O}(M)$}  & \multirow{2}{*}{$\mathcal{O}(MN^2)$} & N Even: $\frac{N^2}{4} {+} 1$ & N Even: $\frac{N^2}{4} {+} 1$ \\ \cline{5-6}
                        &         & & &  N Odd: $\lceil \frac{N^2}{4} \rceil$ & N Odd: $\lceil \frac{N^2}{4} \rceil$ \\ \hline

\multirow{2}{*}{Left Dominance Tree} & Insert: $\mathcal{O}(\log N)$ & \multirow{2}{*}{$\mathcal{O}(M)$}& \multirow{2}{*}{$\mathcal{O}(MN^2)$} & N Even: $\frac{N^2}{4} {+} \frac{N}{2}$  & N Even: $\frac{N^2}{4} {+} \frac{N}{2}$  \\ \cline{2-2} \cline{5-6}
  & Delete: $\mathcal{O}(1)$ & & & N Odd: $\frac{N^2}{4} {+} \frac{N}{2} + \frac{1}{4}$  & N Odd: $\frac{N^2}{4} {+} \frac{N}{2} + \frac{1}{4}$  \\ \hline
  
\multirow{2}{*}{Right Dominance Tree} & Insert: $\mathcal{O}(\log N)$ & \multirow{2}{*}{$\mathcal{O}(M)$}  & \multirow{2}{*}{$\mathcal{O}(MN^2)$} & N Even: $\frac{N^2}{4} {+} 1$  & N Even: $\frac{N^2}{4} {+} 1$  \\ \cline{2-2} \cline{5-6}
  & Delete: $\mathcal{O}(1)$ & & & N Odd: $\lceil \frac{N^2}{4} \rceil$  & N Odd: $\lceil \frac{N^2}{4} \rceil$  \\ \hline

\end{tabular}
\caption{Space and Time complexities of the approaches when considered for non-domination level update problem. $^*$ shows that the worst case space complexity for insertion/deletion are same.}
\label{tab:comparison}
\end{table*}

Table \ref{tab:comparison} shows the comparison among different approaches when considered for non-domination level update problem. Most of the approach perform complete sorting algorithm when either an insertion or a deletion occurs. This table also shows the best and worst case time complexity in case of insertion/deletion. This table also shows the maximum number of comparison in case of insertion/deletion. The worst case space complexity of all the approaches are also shown.

\section{Sorting}\label{sec:sorting}
In this section, we will discuss our proposed non-dominating sorting algorithm. For this purpose, we can use either linear or dominance binary search tree based approach discussed in above section. The process of sorting is described in Algorithm \ref{alg:sorting}. This sorting algorithm is incremental in nature which means this algorithm does not require all the solutions beforehand. This algorithm sorts the solutions as they arrive. So this algorithm can also be used as on-line algorithm \cite{karp1992line}, \cite{albers2003online} because it sorts the solutions as they arrive. 

In this algorithm, initially first solution is inserted into the front and there was no solution in the front so the first solution to be added directly to the front. This solution alone is sorted. Then second solution is added to the front. After the insertion of the second solution, the two solutions are in sorted from. After this the third solution is being inserted in the sorted front and this same process continues for all the solutions.

\noindent
\textbf{Competitive Ratio:} The performance of an on-line algorithm algorithm is evaluated by Competitive Ratio \cite{sleator1985amortized}. As defined in \cite{sleator1985amortized}, the competitive ratio of an on-line algorithm over all possible input sequence is the ratio between the cost incurred by on-line algorithm and the cost incurred by optimal off-line algorithm. Thus an optimal on-line algorithm is one whose competitive ratio is less. An on-line algorithm is known to be competitive if its competitive ratio is bounded.  

Let the on-line algorithm be $\textit{ONSort}$ and the corresponding optimal off-line algorithm be $\textit{OFFSort}$. Let the sequence of solutions to be sorted is $\mathbb{S}$. The cost incurred using $\textit{ONSort}$ is $\textit{ONSort}(\mathbb{S})$ and using $\textit{OFFSort}$ is $\textit{OFFSort}(\mathbb{S})$. Algorithm $\textit{ONSort}$ is known to be $k$-competitive if there exists a constant $c$ such that $\textit{ONSort}(\mathbb{S}) \leq k.\textit{OFFSort}(\mathbb{S}) + c$ for all sequence of solution. Also there should be no relation between the $c$ and the input sequence $\mathbb{S}$.

There are various off-line algorithm proposed for non-dominating sorting e.g. Fast Non-dominated Sort \cite{deb2002fast}, Climbing Sort \cite{mcclymont2012deductive}, Deductive Sort \cite{mcclymont2012deductive}, Arena's Sort \cite{tang2008fast}, ENS-SS \cite{zhang2015efficient}, ENS-BS \cite{zhang2015efficient}. In the worst case, the time complexity of each algorithm is $\mathcal{O}(MN^2)$. Thus the worst case complexity of optimal off-line algorithm for non-dominated sorting is $\mathcal{O}(MN^2)$. The worst case time complexity of the on-line sorting algorithm is given by $\mathcal{O}(MN^3)$ because the insertion of single solution takes $\mathcal{O}(MN^2)$ time and there are $N$ such solutions. Thus the competitive ratio of the on-line algorithm for sorting is given by $N$. The following relation holds $\textit{ONSort}(\mathbb{S}) \leq N.\textit{OFFSort}(\mathbb{S})$

\begin{algorithm}
\begin{algorithmic}[1]
\renewcommand{\algorithmicrequire}{\textbf{Input:}}
\REQUIRE \textit{Population}: $P$
\renewcommand{\algorithmicensure}{\textbf{Output:}}
\ENSURE $\mathcal{F}:$ All non-dominated fronts of $P$ in increasing order of their ranks
\STATE $\mathcal{F} \leftarrow \textit{null}$
\FOR{\textbf{each} $\textit{sol} \in P$}
	\IF{$\mathcal{F} = \textit{null}$}
		\STATE All $\textit{sol}$ to $\mathcal{F}$
	\ELSE	
		\STATE $\textit{InsertLinear}(\mathcal{F},\textit{sol})$
	\ENDIF	
\ENDFOR
\end{algorithmic}
\caption{Sorting$(P)$}
\label{alg:sorting}
\end{algorithm}

\section{Conclusion \& Future Work}\label{sec:conclusion}
In this paper we have proposed the modified version of ENLU approach which is efficient in terms of space. In this paper we have also proposed the new approach based on dominance tree based technique to solve Non-domination Level Update problem. Two variants of this tree are discussed and the update problem can be solved using both the types of tree. This technique inserts the new solution to its correct position and update the dominance level of those solutions which are to be updated. The technique to delete inferior solution is also described. To identify the correct position of the deleted solution dominance tree based approach is used. The maximum number of possible comparisons required in either inserting a solution in the set of fronts or deleting a solution is also obtained. The behaviour of the approach for some special cases are also analysed. At the end, using the proposed technique for Non-domination Level Update problem, a sorting algorithm is provided which does not require all the solutions beforehand unlike all other existing algorithm \cite{deb2002fast}, \cite{mcclymont2012deductive}, \cite{zhang2015efficient}. So the proposed sorting algorithm can be used where all the solutions are not known in advance. This algorithm is on-line so the competitive ratio of this algorithm is proven to be $N$. 

In future we would like to minimize the number of dominance comparison in the situation when a solution is being either inserted or deleted. In this paper we have used the tree structure for the set of fronts and the solutions inside the fronts are considered in linear manner. It would be interesting to see whether the tree structure in the fronts can improve the number of dominance comparison as done in \cite{yakupov2015incremental}, \cite{buzdalov2015various}.

\bibliographystyle{ieeetran}
%\bibliography{mybibfile}
 %Generated by IEEEtran.bst, version: 1.12 (2007/01/11)

\begin{appendices}

\section{Linear Approach}\label{appendix:linear}
The maximum number of comparisons in case of linear approach is given by
\begin{multline*}
f_{\textit{linear}} = n_1 + \left[ \left( n_1{-}1 \right).n_2 + \left( n_2{-}1 \right).n_3 + \ldots  \right.\\ 
                          \left.  + \left( n_{K{-}1}{-}1 \right).n_K  \right]
\end{multline*}
Now our aim is to obtain the maximum value of $f_{\textit{linear}}$ so that the maximum number of dominance comparisons can be obtained. We obtain this in following manner.

\subsection{Number of fronts is 2}\label{app1subsec:nf2}
The population $\mathbb{P}$ of size $N$ is divided in two fronts i.e. $\mathcal{F} = \left\lbrace F_1,F_2 \right\rbrace$. Let $\vert F_1 \vert = n_1$ and $\vert F_2 \vert = n_2$ so $N=n_1+n_2$.
\begin{equation*}
\begin{split}
f_{\textit{linear}}  &= n_1 + \left( n_1{-}1 \right)n_2 \\
  &= n_1 + \left( n_1{-}1 \right) \left( N {-} n_1 \right) \\
  &= n_1N + 2n_1 - n_1^2 - N \\
\frac{df_{\textit{linear}}}{dn_1} &= N + 2 - 2n_1 \\
\frac{d^2f_{\textit{linear}}}{dn_1^2}& = -2
\end{split}
\end{equation*}
\noindent
The maximum value of $f_{\textit{linear}}$ is achieved when $\frac{df_{\textit{linear}}}{dn_1} = 0$ and $\frac{d^2f_{\textit{linear}}}{dn_1^2} < 0$. Thus using this equality we get, $n_1 = \frac{N}{2}+1$ and $n_2 = \frac{N}{2}-1$. The maximum value of $f_{\textit{linear}}$ is as follows:
\begin{equation*}
\begin{split}
f_{\textit{linear}} &= \left( \frac{N}{2}+1 \right)  + \left[ \left( \frac{N}{2}+1 \right)-1 \right] \left( \frac{N}{2}-1 \right) \\
  &= \frac{N^2}{4} + 1
\end{split}
\end{equation*}

\subsection{Number of fronts is 3}\label{app1subsec:nf3}
The population $\mathbb{P}$ of size $N$ is divided in three fronts i.e. $\mathcal{F} = \left\lbrace F_1,F_2,F_3 \right\rbrace$. Let $\vert F_1 \vert = n_1$, $\vert F_2 \vert = n_2$ and $\vert F_3 \vert = n_3$ so $N=n_1+n_2+n_3$.
\begin{equation*}
\begin{split}
f_{\textit{linear}} &= n_1 + \left( n_1{-}1 \right)n_2 + \left( n_2{-}1 \right)n_3 \\
  &= n_1n_2 + n_2n_3 + 2n_1 - N \\
  &= \left( N{-}n_2{-}n_3 \right)n_2 + n_2n_3 + 2\left( N{-}n_2{-}n_3 \right) - N \\
  &= Nn_2 + N - n_2^2 - 2n_2 - 2n_3 
\end{split}
\end{equation*}

\noindent
The value of $f_{\textit{linear}}$ will be maximized when $n_3 = 0$. Thus $N = n_1 + n_2$. From the subsection \ref{app1subsec:nf2}, we can conclude that when the population is divided into $2$ fronts then the maximum number of dominance comparison will be $\frac{N^2}{4} + 1$. Hence for maximum number of dominance comparison, $n_1=\frac{N}{2}+1,n_2=\frac{N}{2}-1,n_3=0$.

\subsection{Number of fronts is 4}\label{app1subsec:nf4}
The population $\mathbb{P}$ of size $N$ is divided in four fronts i.e. $\mathcal{F} = \left\lbrace F_1,F_2,F_3,F_4 \right\rbrace$. $\vert F_1 \vert = n_1$, $\vert F_2 \vert = n_2$, $\vert F_3 \vert = n_3$ and $\vert F_4 \vert = n_4$ so $N=n_1+n_2+n_3+n_4$.
\begin{equation*}
\begin{split}
f_{\textit{linear}} &= n_1 + \left( n_1{-}1 \right)n_2 + \left( n_2{-}1 \right)n_3 + \left( n_3{-}1 \right)n_4 \\
  &= n_1n_2 + n_2n_3 + n_3n_4 + 2n_1 - N \\
  &= n_2 \left( n_1+n_3 \right) + n_3n_4 + 2n_1 - N \\
  &= n_2 \left( N {-} n_2 {-} n_4 \right) + n_3\left( N{-}n_1{-}n_2{-}n_3 \right) +2n_1 - N \\  
  &= \left( n_2+n_3 \right)N - \left( n_2^2+n_3^2 \right) - \left( n_1+n_2 \right)n_3 + 2n_1 - \\
  & \mathrel{\phantom{=}} n_2n_4 - N
\end{split}
\end{equation*}

The value of $f_{\textit{linear}}$ will be maximized when $n_4 = 0$. Thus $N = n_1 + n_2 + n_3$. From the subsection \ref{app1subsec:nf3}, we can conclude that when the population is divided into $3$ fronts then the maximum value of the function is $\frac{N^2}{4} + 1$. Thus For maximum number of dominance comparison, $n_1=\frac{N}{2}+1,n_4=\frac{N}{2}-1,n_3=0,n_4=0$.

\subsection{Number of fronts is 5}\label{app1subsec:nf5}
The population $\mathbb{P}$ of size $N$ is divided in five fronts i.e $\mathcal{F} = \left\lbrace F_1,F_2,F_3,F_4,F_5 \right\rbrace$. $\vert F_1 \vert = n_1$, $\vert F_2 \vert = n_2$, $\vert F_3 \vert = n_3$, $\vert F_4 \vert = n_4$ and $\vert F_5 \vert = n_5$ so $N=n_1+n_2+n_3+n_4+n_5$.
\begin{equation*}
\begin{split}
f_{\textit{linear}} &= n_1 + \left( n_1{-}1 \right)n_2 + \left( n_2{-}1 \right)n_3 + \left( n_3{-}1 \right)n_4 + \left( n_4{-}1 \right)n_5 \\
  &= n_1n_2 + n_2n_3 + n_3n_4 +n_4n_5 + 2n_1 - N \\
  &= n_1n_2 + n_2 \left( N {-} n_1 {-} n_2 {-} n_4 {-} n_5 \right) + \left( N {-} n_1 {-} n_2 {-} n_4 {-} \right.  \\
  & \mathrel{\phantom{=}} \left. n_5 \right) n_4 + n_4 n_5 + 2n_1 - N \\
  &= \left( n_2+n_4 \right)N - \left( n_2^2 + n_4^2 \right) - 2n_2n_4  - n_1n_4 + 2n_1 -  \\
  & \mathrel{\phantom{=}} N - n_2n_5
\end{split}
\end{equation*}

\noindent
The value of $f_{\textit{linear}}$ will be maximized when $n_5 = 0$. Thus $N = n_1 + n_2 + n_3 + n_4$. From the subsection \ref{app1subsec:nf4}, we can conclude that when the population is divided into $4$ fronts then the maximum value of the function is $\frac{N^2}{4} + 1$. Hence For maximum number of dominance comparison, $n_1=\frac{N}{2}+1,n_2=\frac{N}{2}-1,n_3=0,n_4=0,n_5=0$.

Thus we can conclude that the maximum number of comparisons in case of linear approach is  $\frac{N^2}{4} + 1$ which occurs when there are two fronts i.e. $\mathcal{F} = \left\lbrace F_1,F_2 \right\rbrace$. The cardinality of each front is $\vert F_1 \vert = \frac{N}{2}+1$ and $\vert F_2 \vert = \frac{N}{2}-1$.

\section{Left Dominance Binary Search Tree Based Approach}\label{appendix:left}
The maximum number of comparisons in case of left dominance binary search tree based approach is given by
\begin{equation*}
\begin{split}
f_{\textit{ltree}}  &= \left[ n_{\lceil \frac{\textit{mid}}{2^0} \rceil} {+} n_{\lceil \frac{\textit{mid}}{2^1} \rceil} {+} \ldots {+} n_{{\lceil \frac{\textit{mid}}{2^{h}} \rceil}} \right] {+} \left[ \left( n_1{-}1 \right).n_2 \right.  \\ 
& \mathrel{\phantom{=}} \left.  + \left( n_2{-}1 \right).n_3 + \ldots + \left( n_{K{-}1}{-}1 \right).n_K  \right] \\
&= \alpha + \beta 
\end{split}
\end{equation*}
For maximum value of the $f_{\textit{ltree}}$, $\alpha$ and $\beta$ both should be maximized. The maximum value of $\alpha$ can be $N$. So we focus on maximizing $\beta$.
\begin{equation*}
\beta  =  \left( n_1{-}1 \right).n_2 + \left( n_2{-}1 \right).n_3 + \ldots + \left( n_{K{-}1}{-}1 \right).n_K 
\end{equation*} 
Now our aim is to obtain the maximum value of $\beta$ so that the maximum number of dominance comparisons can be obtained. We obtain this in following manner.

\subsection{Number of fronts is 2}\label{app2subsec:nflt2}
The population $\mathbb{P}$ of size $N$ is divided in two fronts i.e. $\mathcal{F} = \left\lbrace F_1,F_2 \right\rbrace$. Let $\vert F_1 \vert = n_1$ and $\vert F_2 \vert = n_2$ so $N=n_1+n_2$.
\begin{equation*}
\begin{split}
\beta &= \left( n_1{-}1 \right)n_2 \\
      &= \left( n_1{-}1 \right) \left( N {-} n_1 \right) \\
      &= n_1N + n_1 - n_1^2 - N \\
\frac{df_{\beta}}{dn_1} &= N + 1 - 2n_1 \\
\frac{d^2f_{\beta}}{dn_1^2}& = -2
\end{split}
\end{equation*}
\noindent
The maximum value of $\beta$ is achieved when $\frac{d\beta}{dn_1} = 0$ and $\frac{d^2\beta}{dn_1^2} < 0$. Thus using this equality we get, $n_1 = \frac{N+1}{2}$ and $n_2 = \frac{N-1}{2}$. The maximum value of $\beta$ is as follows:
\begin{equation*}
\begin{split}
f_{\textit{ltree}} &= \left[ \left( \frac{N+1}{2} \right)-1 \right] \left( \frac{N-1}{2} \right) \\
  &= \frac{(N-1)^2}{4}
\end{split}
\end{equation*}

\subsection{Number of fronts is 3}\label{app2subsec:nflt3}
The population $\mathbb{P}$ of size $N$ is divided in three fronts i.e $\mathcal{F} = \left\lbrace F_1,F_2,F_3 \right\rbrace$. Let $\vert F_1 \vert = n_1$, $\vert F_2 \vert = n_2$ and $\vert F_3 \vert = n_3$ so $N=n_1+n_2+n_3$.
\begin{equation*}
\begin{split}
\beta &= \left( n_1{-}1 \right)n_2 + \left( n_2{-}1 \right)n_3 \\
  &= n_1n_2 + n_2n_3 + n_1 - N \\
  &= \left( N{-}n_2{-}n_3 \right)n_2 + n_2n_3 + \left( N{-}n_2{-}n_3 \right) - N \\
  &= Nn_2 - n_2^2 - n_2 - n_3 
\end{split}
\end{equation*}

\noindent
The value of $\beta$ will be maximized when $n_3 = 0$. Thus $N = n_1 + n_2$. From the subsection \ref{app2subsec:nflt2}, we can conclude that when the population is divided into $2$ fronts then the maximum number of dominance comparison will be $\frac{(N-1)^2}{4}$. Thus for maximum number of dominance comparison, $n_1=\frac{N+1}{2},n_2=\frac{N-1}{2},n_3=0$.

\subsection{Number of fronts is 4}\label{app2subsec:nflt4}
The population $\mathbb{P}$ of size $N$ is divided in four fronts i.e. $\mathcal{F} = \left\lbrace F_1,F_2,F_3,F_4 \right\rbrace$. $\vert F_1 \vert = n_1$, $\vert F_2 \vert = n_2$, $\vert F_3 \vert = n_3$ and $\vert F_4 \vert = n_4$ so $N=n_1+n_2+n_3+n_4$.
\begin{equation*}
\begin{split}
\beta &= \left( n_1{-}1 \right)n_2 + \left( n_2{-}1 \right)n_3 + \left( n_3{-}1 \right)n_4 \\
  &= n_1n_2 + n_2n_3 + n_3n_4 + n_1 - N \\
  &= n_2 \left( n_1+n_3 \right) + n_3n_4 + n_1 - N \\
  &= n_2 \left( N {-} n_2 {-} n_4 \right) + n_3\left( N{-}n_1{-}n_2{-}n_3 \right) + n_1 - N \\  
  &= \left( n_2+n_3 \right)N - \left( n_2^2+n_3^2 \right) - \left( n_1+n_2 \right)n_3 + n_1 - \\
  & \mathrel{\phantom{=}} n_2n_4 - N
\end{split}
\end{equation*}

The value of $\beta$ will be maximized when $n_4 = 0$. Thus $N = n_1 + n_2 + n_3$. From the subsection \ref{app2subsec:nflt3}, we can conclude that when the population is divided into $3$ fronts then the maximum value of the function is $\frac{(N-1)^2}{4}$. Hence for maximum number of dominance comparison, $n_1=\frac{N+1}{2},n_2=\frac{N-1}{2},n_3=0,n_4=0$.

\subsection{Number of fronts is 5}\label{app2subsec:nflt5}
The population $\mathbb{P}$ of size $N$ is divided in five fronts i.e. $\mathcal{F} = \left\lbrace F_1,F_2,F_3,F_4,F_5 \right\rbrace$. $\vert F_1 \vert = n_1$, $\vert F_2 \vert = n_2$, $\vert F_3 \vert = n_3$, $\vert F_4 \vert = n_4$ and $\vert F_5 \vert = n_5$ so $N=n_1+n_2+n_3+n_4+n_5$.
\begin{equation*}
\begin{split}
\beta &= \left( n_1{-}1 \right)n_2 + \left( n_2{-}1 \right)n_3 + \left( n_3{-}1 \right)n_4 + \left( n_4{-}1 \right)n_5 \\
  &= n_1n_2 + n_2n_3 + n_3n_4 +n_4n_5 + n_1 - N \\
  &= n_1n_2 + n_2 \left( N {-} n_1 {-} n_2 {-} n_4 {-} n_5 \right) + \left( N {-} n_1 {-} n_2 {-} n_4 {-} \right.  \\
  & \mathrel{\phantom{=}} \left. n_5 \right) n_4 + n_4 n_5 + n_1 - N \\
  &= \left( n_2+n_4 \right)N - \left( n_2^2 + n_4^2 \right) - 2n_2n_4  - n_1n_4 + n_1 -  \\
  & \mathrel{\phantom{=}} N - n_2n_5
\end{split}
\end{equation*}

\noindent
The value of $\beta$ will be maximized when $n_5 = 0$. Thus $N = n_1 + n_2 + n_3 + n_4$. From the subsection \ref{app2subsec:nflt4}, we can conclude that when the population is divided into $4$ fronts then the maximum value of the function is $\frac{(N-1)^2}{4}$. Thus for maximum number of dominance comparison, $n_1=\frac{N+1}{2},n_2=\frac{N-1}{2}-1,n_3=0,n_4=0,n_5=0$.

Thus we can conclude that the maximum value of $\beta$ is $\frac{(N-1)^2}{4}$ which occurs when there are two fronts i.e. $\mathcal{F} = \left\lbrace F_1,F_2 \right\rbrace$. The cardinality of each front is $\vert F_1 \vert = \frac{N+1}{2}$ and $\vert F_2 \vert = \frac{N-1}{2}$. The maximum comparison occurs when the inserted/deleted solution dominate the solution in first front. As the tree is left dominance so in case of two fronts the inserted/deleted solution is compared with both the fronts (first $F_2$ then $F_1$) so the maximum value of $\alpha = n_1 + n_2 = N$. Thus the maximum number of comparison when dominance tree based approach is used is given by 
\begin{equation*}
f_{\textit{ltree}} = \alpha + \beta = N + \frac{(N-1)^2}{4} = \frac{N^2}{4} + \frac{N}{2} + \frac{1}{4}
\end{equation*}

\section{Right Dominance Binary Search Tree Based Approach}\label{appendix:right}
The maximum number of comparisons in case of right dominance binary search tree based approach is given by
\begin{equation*}
\begin{split}
f_{\textit{rtree}}  &= \left[ n_{\lfloor \frac{\textit{mid}}{2^0} \rfloor} {+} n_{\lfloor \frac{\textit{mid}}{2^1} \rfloor} {+} \ldots {+} n_{{\lfloor \frac{\textit{mid}}{2^{h}} \rfloor}} \right] + \left[ \left( n_1{-}1 \right).n_2 \right. \\ 
& \mathrel{\phantom{=}} \left. + \left( n_2{-}1 \right).n_3 + \ldots + \left( n_{K{-}1}{-}1 \right).n_K  \right] \\ 
&= \alpha + \beta 
\end{split}
\end{equation*}

For maximum value of the $f_{\textit{rtree}}$, $\alpha$ and $\beta$ both should be maximized. The maximum value of $\alpha$ can be $N$. So we focus on maximizing $\beta$.

The maximum value of $\beta$ is $\frac{(N-1)^2}{4}$ which occurs when there are two fronts i.e. $\mathcal{F} = \left\lbrace F_1,F_2 \right\rbrace$. The cardinality of each front is $\vert F_1 \vert = \frac{N+1}{2}$ and $\vert F_2 \vert = \frac{N-1}{2}$. The maximum number of comparisons occurs when the inserted/deleted solution dominate the solution in first front. As the tree is right dominance so in case of two fronts the inserted/deleted solution is compared with only first front so the maximum value of $\alpha = n_1 = \frac{N+1}{2}$. Thus the maximum number of comparison when right dominance tree based approach is used is given by 
\begin{equation*}
\begin{split}
f_{\textit{rtree}} &= \alpha + \beta = \left( \frac{N+1}{2} \right) + \frac{(N-1)^2}{4}\\
            &= \frac{N^2+3}{4} = \left\lceil\dfrac{N^2}{4} \right\rceil
\end{split}
\end{equation*}

\end{appendices}

\end{document}